\documentclass
[prb,showpacs,floats,amsmath,amssymb,a4paper,11pt,tightenlines]{revtex4}%
\usepackage{epsf}
\usepackage{graphicx}
\usepackage{amsmath}
\usepackage{amsfonts}
\usepackage{amssymb}
\usepackage{dcolumn}%
\providecommand{\U}[1]{\protect\rule{.1in}{.1in}}

\begin{document}
\title{Theory of spin Hall magnetoresistance}
\date{\today}
\author{Yan-Ting Chen$^{1}$, Saburo Takahashi$^{2}$, Hiroyasu Nakayama$^{2}$, Matthias
Althammer$^{3,4}$, Sebastian T. B. Goennenwein$^{3}$, Eiji Saitoh$^{2,5,6,7}$,
and Gerrit E. W. Bauer$^{2,5,1}$}
\affiliation{$^{1}$ Kavli Institute of NanoScience, Delft University of Technology,
Lorentzweg 1, 2628 CJ Delft, The Netherlands }
\affiliation{$^{2}$ Institute for Materials Research, Tohoku University, Sendai, Miyagi
980-8577, Japan}
\affiliation{$^{3}$ Walther-Mei{\ss }ner-Institut, Bayerische Akademie der Wissenschaften,
85748 Garching, Germany}
\affiliation{$^{4}$ University of Alabama, Center for Materials for Information Technology
MINT, Dept Chem, Tuscaloosa, AL 35487, USA}
\affiliation{$^{5}$ WPI Advanced Institute for Materials Research, Tohoku University,
Sendai 980-8577, Japan}
\affiliation{$^{6}$ CREST, Japan Science and Technology Agency, Sanbancho, Tokyo 102-0075, Japan}
\affiliation{$^{7}$ The Advanced Science Research Center, Japan Atomic Energy Agency, Tokai
319-1195, Japan}

\begin{abstract}
We present a theory of the spin Hall magnetoresistance (SMR) in multilayers
made from an insulating ferromagnet F, such as yttrium iron garnet (YIG), and
a normal metal N with spin-orbit interactions, such as platinum (Pt). The SMR
is induced by the simultaneous action of spin Hall and inverse spin Hall
effects and therefore a non-equilibrium proximity phenomenon. We compute the
SMR in F$\vert$N and F$\vert$N$\vert$F layered systems, treating N by spin-diffusion
theory with quantum mechanical boundary conditions at the interfaces in terms
of the spin-mixing conductance. Our results explain the experimentally
observed spin Hall magnetoresistance in N$\vert$F bilayers. For F$\vert$N$\vert$F spin
valves we predict an enhanced SMR amplitude when magnetizations are collinear.
The SMR and the spin-transfer torques in these trilayers can be controlled by
the magnetic configuration.

\end{abstract}

\maketitle

\section{Introduction}

Spin currents are a central theme in spintronics since they are intimately
associated with the manipulation and transport of spins in small structures
and devices.\cite{Bader10,Sinova12} Spin currents can be generated by means of
the spin Hall effect (SHE) and detected by the inverse spin Hall effect
(ISHE).\cite{Jungwirth12} Of special interest are multilayers made of normal
metals (N) and ferromagnets (F). When an electric current flows through N, an
SHE spin current flows towards the interfaces, where it can be absorbed as a
spin-transfer torque (STT) on the ferromagnet. This STT affects the
magnetization damping\cite{Ando08} or even switches the
magnetization.\cite{Miron11,Liu12} The ISHE can be used to detect spin
currents pumped by the magnetization dynamics excited by
microwaves\cite{Saitoh06,Mosendz10-1,Mosendz10-2,Czeschka11} or temperature
gradients (spin Seebeck effect).\cite{Uchida08,Jaworski10}

Recently, magnetic insulators have attracted the attention of the spintronics
community. Yttrium iron garnets (YIG), a class of ferrimagnetic insulators
with a large band gap, are interesting because of their very low magnetization
damping. Their magnetization can be activated thermally to generate the spin
Seebeck effect in YIG$|$Pt bilayers.\cite{Uchida10,Weiler12} By means of the
SHE, spin waves can be electrically excited in YIG via a Pt contact, and, via
the ISHE, subsequently detected electrically in another Pt
contact.\cite{Kajiwara10} Spin transport at an N$|$F interface is governed by
the complex spin-mixing conductance $G_{\uparrow\downarrow}$.\cite{Brataas00}
The prediction of a large real part of $G_{\uparrow\downarrow}$ for interfaces
of YIG with simple metals by first principles calculations\cite{Jia11} has been
confirmed by experiments.\cite{Heinrich11}

Magnetoresistance (MR) is the property of a material to change the value of
its electrical resistance under an external magnetic field. In normal metals
its origin is the Lorentz force.\cite{Ashcroft76} The dependence of the
resistance on the angle between current and magnetization in metallic
ferromagnets is called anisotropic magnetoresistance (AMR). The transverse
component of the AMR is also called the planar Hall effect (PHE),
\textit{i.e.} the transverse (Hall) voltage found in ferromagnets when the
magnetization is rotated in the plane of the film.\cite{McGuire75,Thompson75}
Both effects are symmetric with respect to magnetization reversal, which
distinguishes them from the anomalous Hall effect (AHE) for magnetizations
normal to the film, which changes sign under magnetization
reversal.\cite{Nagaosa10} The physical origin of AMR, PHE, and AHE is the
spin-orbit interaction, in contrast to the giant magnetoresistance (GMR),
which reflects the change in resistance that accompanies the magnetic
field-induced magnetic configuration in magnetic multilayers.\cite{Fert08}

Here we propose a theory for a recently discovered magnetoresistance effect in
Pt$|$YIG bilayer systems.\cite{Weiler12,Huang12,Nakayama12} This MR is
remarkable since YIG\ is a very good electric insulator such that a charge
current can only flow in Pt. We explain this unusual magnetoresistance not in
terms of an equilibrium static magnetic proximity polarization in
Pt,\cite{Huang12} but rather in terms of a non-equilibrium proximity effect
caused by the simultaneous action of the SHE and ISHE and therefore call it
spin Hall magnetoresistance (SMR). This effect scales like the square of the
spin Hall angle and is modulated by the magnetization direction in YIG via the
spin-transfer at the N$|$F interface. Our explanation is similar to the Hanle
effect-induced magnetoresistance in the two-dimensional electron gas proposed
by Dyakonov.\cite{Dyakonov07} Here we present the details of our theory, which
is based on the spin-diffusion approximation in the N layer in the presence of
spin-orbit interactions\cite{Takahashi06} and quantum mechanical boundary
conditions at the interface in terms of the spin-mixing
conductance.\cite{Brataas00,Jia11} We also address F$|$N$|$F spin valves with
electric currents applied parallel to the interface(s) with the additional
degree of freedom of the relative angle between the two magnetizations directions.

\begin{figure}[ptb]
\includegraphics[width=0.5\textwidth,angle=0]{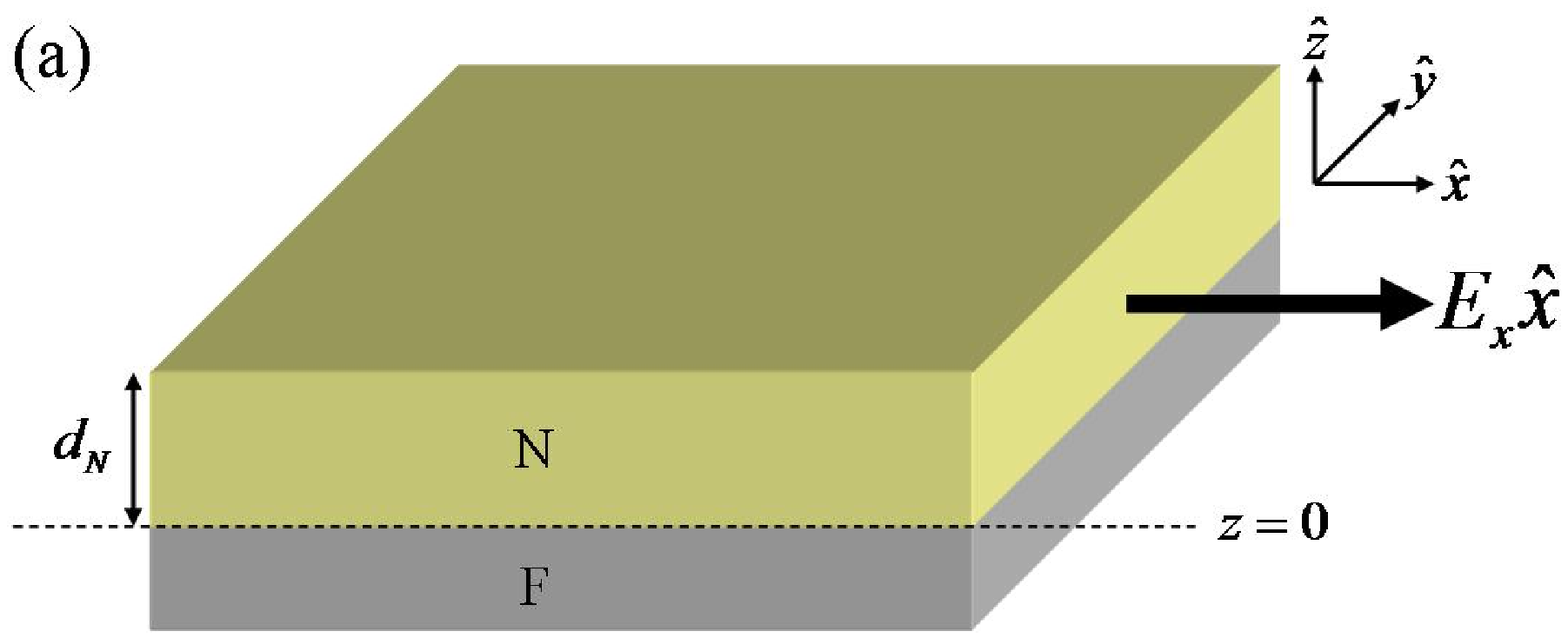}
\includegraphics[width=0.5\textwidth,angle=0]{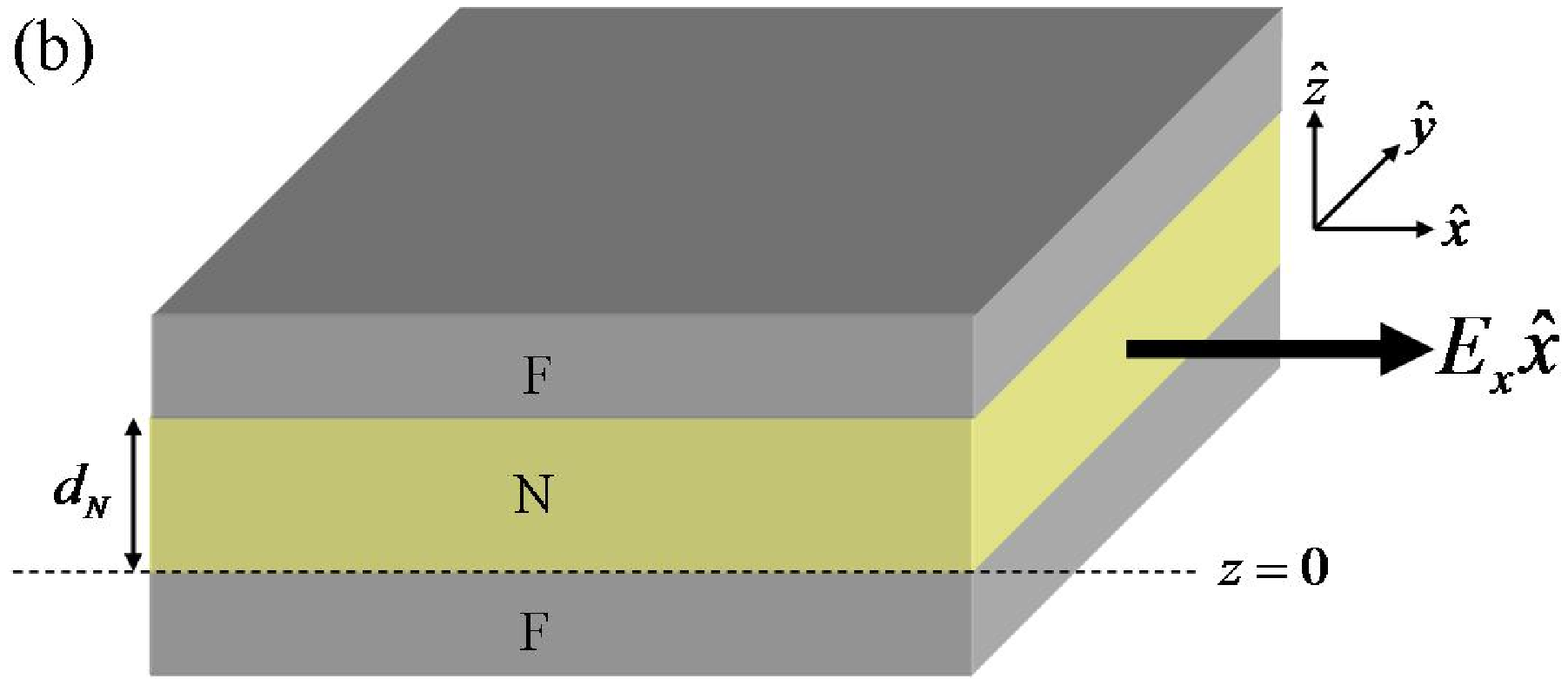}\caption{(a)
N$|$F bilayer and (b) F$|$N$|$F trilayer systems considered here, where F is a
ferromagnetic insulator and N a normal metal.}%
\label{bilayer-trilayer}%
\end{figure}

This paper is organized as follows. We present the model, \textit{i.e.}
spin-diffusion with proper boundary conditions in Sec.~\ref{general method}.
In Sec.~\ref{vac-pt-yig}, we consider an N$|$F bilayer as shown in
Fig.~\ref{bilayer-trilayer}~(a). We obtain spin accumulation, spin currents
and finally the measured charge currents that are compared with the
experimental SMR. We also find and discuss that the imaginary part of the
spin-mixing conductance generates an AHE. F$|$N$|$F
(Fig.~\ref{bilayer-trilayer}~(b)) spin valves are investigated in
Sec.~\ref{yig-pt-yig}, which show an enhanced SMR for spacers thinner than the
spin-flip diffusion length. We summarize the results and give conclusions in
Sec.~\ref{summary}.

\section{Transport theory in metals in contact with a magnetic insulator}

\label{general method} The spin current density in the non-relativistic limit
\begin{equation}
\overleftrightarrow{\mathbf{j}_{s}}=en\left\langle \vec{v}\otimes
\boldsymbol{\vec{\sigma}}+\boldsymbol{\vec{\sigma}}\otimes\vec{v}\right\rangle
/2=\left(  \vec{j}_{sx},\vec{j}_{sy},\vec{j}_{sz}\right)  ^{T}=\left(  \vec
{j}_{s}^{x},\vec{j}_{s}^{y},\vec{j}_{s}^{z}\right)  \label{sc}%
\end{equation}
is a second-order tensor (in units of the charge current density $\vec{j}%
_{c}=en\left\langle \vec{v}\right\rangle $), where $e=|e|$ is the electron
charge, $n$ is the density of the electrons, $\vec{v}$ is the velocity
operator, $\boldsymbol{\vec{\sigma}}$ is the vector of Pauli spin matrices,
and $\left\langle \cdots\right\rangle $ denotes an expectation value. The row
vectors $\vec{j}_{si}=en\left\langle \vec{v}\boldsymbol{\sigma}_{i}%
+\boldsymbol{\sigma}_{i}\vec{v}\right\rangle /2$ in Eq.~(\ref{sc}) are the
spin current densities polarized in the $\hat{\imath}$-direction, while the
column vectors $\vec{j}_{s}^{j}=en\left\langle v_{j}\boldsymbol{\vec{\sigma}%
}+\boldsymbol{\vec{\sigma}}v_{j}\right\rangle /2$ denote the spin current
densities with polarization $\boldsymbol{\vec{\sigma}}$ flowing in the
$\hat{\jmath}$-direction. Ohm's law for metals with spin-orbit interactions
can be summarized by the relation between thermodynamic driving forces and
currents that reflects Onsager's reciprocity by the symmetry of the response
matrix:\cite{Takahashi06}
\begin{equation}
\left(
\begin{array}
[c]{c}%
\vec{j}_{c}\\
\vec{j}_{sx}\\
\vec{j}_{sy}\\
\vec{j}_{sz}%
\end{array}
\right)  =\sigma\left(
\begin{array}
[c]{cccc}%
1 & \theta_{\mathrm{SH}}\hat{x}\times & \theta_{\mathrm{SH}}\hat{y}\times &
\theta_{\mathrm{SH}}\hat{z}\times\\
\theta_{\mathrm{SH}}\hat{x}\times & 1 & 0 & 0\\
\theta_{\mathrm{SH}}\hat{y}\times & 0 & 1 & 0\\
\theta_{\mathrm{SH}}\hat{z}\times & 0 & 0 & 1
\end{array}
\right)  \left(
\begin{array}
[c]{c}%
-\vec{\nabla}\mu_{0}/e\\
-\vec{\nabla}\mu_{sx}/(2e)\\
-\vec{\nabla}\mu_{sy}/(2e)\\
-\vec{\nabla}\mu_{sz}/(2e)
\end{array}
\right)  , \label{response}%
\end{equation}
where $\vec{\mu}_{s}=\left(  \mu_{sx},\mu_{sy},\mu_{sz}\right)  ^{T}-\mu
_{0}\hat{1}$ is the spin accumulation, \textit{i.e.} the spin-dependent
chemical potential relative to the charge chemical potential $\mu_{0}$,
$\sigma$ is the electric conductivity, $\theta_{\mathrm{SH}}$ is the spin Hall
angle, and \textquotedblleft$\times$\textquotedblright\ denotes the vector
cross product operating on the gradients of the spin-dependent chemical
potentials. The spin Hall effect is represented by the lower non-diagonal
elements that generate the spin currents in the presence of an applied
electric field, in the following chosen to be in the $\hat{x}$-direction
$\vec{E}=E_{x}\hat{x}=-\hat{x}\partial_{x}\mu_{0}/e$. The inverse spin Hall
effect is governed by elements above the diagonal that connect the gradients
of the spin accumulations to the charge current density.

The spin accumulation $\vec{\mu}_{s}$ is obtained from the spin-diffusion
equation in the normal metal%
\begin{equation}
\nabla^{2}\vec{\mu}_{s}=\frac{\vec{\mu}_{s}}{\lambda^{2}},
\end{equation}
where the spin-diffusion length $\lambda=\sqrt{D\tau_{\mathrm{sf}}}$ is
expressed in terms of the charge diffusion constant $D$ and spin-flip
relaxation time $\tau_{\mathrm{sf}}$.\cite{Valet93} For films with thickness
$d_{N}$ in the $\hat{z}$-direction
\begin{equation}
\vec{\mu}_{s}\left(  z\right)  =\vec{A}e^{-z/\lambda}+\vec{B}e^{z/\lambda},
\label{mus_0}%
\end{equation}
where the constant column vectors $\vec{A}$ and $\vec{B}$ are determined by
the boundary conditions at the interfaces.

According to Eq. (\ref{response}), the spin current in N consists of diffusion
and spin Hall drift contributions. Since we are considering a system
homogeneous in the $x$-$y$ plane, we focus on the spin current density flowing
in the $\hat{z}$-direction
\begin{equation}
\vec{j}_{s}^{z}(z)=-\frac{\sigma}{2e}\partial_{z}\vec{\mu}_{s}-j_{s0}%
^{\mathrm{SH}}\hat{y}, \label{js-flows-z}%
\end{equation}
where $j_{s0}^{\mathrm{SH}}=\theta_{\mathrm{SH}}\sigma E_{x}$ is the bare
spin Hall current, \textit{i.e.}, the spin current generated directly by the SHE.

The boundary conditions require that $\vec{j}_{s}^{z}(z)$ is continuous at the
interfaces $z=d_{N}$ and $z=0$. The spin current at a vacuum (V) interface
vanishes, $\vec{j}_{s}^{(\mathrm{V})}=0$. The spin current density $\vec
{j}_{s}^{(\mathrm{F})}$ at a magnetic interface is governed by the spin
accumulation and spin-mixing conductance:\cite{Brataas00}
\begin{equation}
e\vec{j}_{s}^{\left(  \mathrm{F}\right)  }\left(  \hat{m}\right)  =G_{r}%
\hat{m}\times\left(  \hat{m}\times\vec{\mu}_{s}\right)  +G_{i}\left(  \hat
{m}\times\vec{\mu}_{s}\right)  , \label{js_interface}%
\end{equation}
where $\hat{m}=\left(  m_{x},m_{y},m_{z}\right)  ^{T}$ is a unit vector along
the magnetization and $G_{\uparrow\downarrow}=G_{r}+iG_{i}$ the complex
spin-mixing interface conductance per unit area. The imaginary part $G_{i}$
can be interpreted as an effective exchange field acting on the spin
accumulation. A positive current in Eq.~(\ref{js_interface}) corresponds to
up-spins flowing from F towards N. Since F is an insulator, this spin current
density is proportional to the spin-transfer acting on the ferromagnet
\begin{equation}
\vec{\tau}_{\mathrm{stt}}=-\frac{\hbar}{2e}\hat{m}\times\left(  \hat{m}%
\times\vec{j}_{s}^{(\mathrm{F})}\right)  =\frac{\hbar}{2e}\vec{j}%
_{s}^{(\mathrm{F})} \label{stt}%
\end{equation}

With these boundary conditions we determine the coefficients $\vec{A}$ and
$\vec{B}$, which leads to the spin accumulation
\begin{equation}
\vec{\mu}_{s}=\frac{2e\lambda}{\sigma}\left[  -\left(  j_{s0}^{\mathrm{SH}%
}\hat{y}+\vec{j}_{s}^{z}(d_{N})\right)  \cosh\frac{z}{\lambda}+\left(
j_{s0}^{\mathrm{SH}}\hat{y}+\vec{j}_{s}^{(\mathrm{F})}\left(  \hat{m}\right)
\right)  \cosh\frac{z-d_{N}}{\lambda}\right]  /\sinh\frac{d_{N}}{\lambda},
\label{mu}%
\end{equation}
where $\vec{j}_{s}^{z}(d_{N})=0$ for F$\left(  \hat{m}\right)  |$N$|$V
bilayers and $\vec{j}_{s}^{z}(d_{N})=-\vec{j}_{s}^{(\mathrm{F})}\left(
\hat{m}^{\prime}\right)  $ for F$\left(  \hat{m}\right)  |$N$|$F$\left(
\hat{m}^{\prime}\right)  $ spin valves.

\section{N$\vert$F bilayers}

\label{vac-pt-yig}

In the bilayer the spin accumulation (\ref{mu}) is
\begin{equation}
\vec{\mu}_{s}(z)=  -\hat{y}\mu_{s}^{0}\frac{\sinh\frac{2z-d_{N}}{2\lambda
}}{\sinh\frac{d_{N}}{2\lambda}}+\vec{j}_{s}^{\left(  \mathrm{F}\right)  }\left(  \hat{m}\right)
\frac{2e\lambda}{\sigma}\frac{\cosh\frac{z-d_{N}}{\lambda}}{\sinh\frac
{d_{N}}{\lambda}},
\end{equation}
where $\mu_{s}^{0}\equiv\left\vert \vec{\mu}_{s}(0)\right\vert =(2e\lambda
/\sigma)j_{s0}^{\mathrm{SH}}\tanh\left[  d_{N}/\left(  2\lambda\right)
\right]  $ is the spin accumulation at the interface in the absence of
spin-transfer, \textit{i.e.}, when $G_{\uparrow\downarrow}=0$.

Using Eq.~(\ref{js_interface}), the spin accumulation at $z=0$ becomes
\begin{equation}
\vec{\mu}_{s}(0)=\hat{y}\mu_{s}^{0}+\frac{2\lambda}{\sigma}\left\{
G_{r}\left[  \hat{m}\left(  \hat{m}\cdot\vec{\mu}_{s}(0)\right)  -\vec{\mu
}_{s}(0)\right]  +G_{i}\hat{m}\times\vec{\mu}_{s}(0)\right\}  \coth\frac
{d_{N}}{\lambda}. \label{mus0_1}%
\end{equation}
With
\begin{align}
\hat{m}\cdot\vec{\mu}_{s}(0)  &  =m_{y}\mu_{s}^{0},\label{m_dot_mus0}\\
\hat{m}\times\vec{\mu}_{s}(0)  &  =\mu_{s}^{0}\frac{\sigma\hat{m}\times\hat
{y}+\hat{m}m_{y}2\lambda G_{i}\coth\frac{d_{N}}{\lambda}}{\sigma+2\lambda
G_{r}\coth\frac{d_{N}}{\lambda}}-\vec{\mu}_{s}(0)\frac{2\lambda G_{i}%
\coth\frac{d_{N}}{\lambda}}{\sigma+2\lambda G_{r}\coth\frac{d_{N}}{\lambda}},
\label{m_cross_mus0}%
\end{align}%
\begin{align}
\vec{\mu}_{s}(0)  &  =\hat{y}\mu_{s}^{0}\frac{1+\frac{2\lambda}{\sigma}%
G_{r}\coth\frac{d_{N}}{\lambda}}{\left(  1+\frac{2\lambda}{\sigma}G_{r}%
\coth\frac{d_{N}}{\lambda}\right)  ^{2}+\left(  \frac{2\lambda}{\sigma}%
G_{i}\coth\frac{d_{N}}{\lambda}\right)  ^{2}}\nonumber\\
&  +\hat{m}m_{y}\mu_{s}^{0}\frac{\frac{2\lambda}{\sigma}G_{r}\coth\frac{d_{N}%
}{\lambda}\left(  1+\frac{2\lambda}{\sigma}G_{r}\coth\frac{d_{N}}{\lambda
}\right)  +\left(  \frac{2\lambda}{\sigma}G_{i}\coth\frac{d_{N}}{\lambda
}\right)  ^{2}}{\left(  1+\frac{2\lambda}{\sigma}G_{r}\coth\frac{d_{N}%
}{\lambda}\right)  ^{2}+\left(  \frac{2\lambda}{\sigma}G_{i}\coth\frac{d_{N}%
}{\lambda}\right)  ^{2}}\nonumber\\
&  +\left(  \hat{m}\times\hat{y}\right)  \mu_{s}^{0}\frac{\frac{2\lambda
}{\sigma}G_{i}\coth\frac{d_{N}}{\lambda}}{\left(  1+\frac{2\lambda}{\sigma
}G_{r}\coth\frac{d_{N}}{\lambda}\right)  ^{2}+\left(  \frac{2\lambda}{\sigma
}G_{i}\coth\frac{d_{N}}{\lambda}\right)  ^{2}},
\end{align}
the spin current through the F$|$N interface then reads
\begin{equation}
\vec{j}_{s}^{(\mathrm{F})}=\frac{\mu_{s}^{0}}{e}\hat{m}\times\left(  \hat
{m}\times\hat{y}\right)  \sigma\operatorname{Re}\frac{G_{\uparrow\downarrow}%
}{\sigma+2\lambda G_{\uparrow\downarrow}\coth\frac{d_{N}}{\lambda}}+\frac
{\mu_{s}^{0}}{e}\left(  \hat{m}\times\hat{y}\right)  \sigma\operatorname{Im}%
\frac{G_{\uparrow\downarrow}}{\sigma+2\lambda G_{\uparrow\downarrow}\coth
\frac{d_{N}}{\lambda}}.
\end{equation}
The spin accumulation
\begin{equation}
\frac{\vec{\mu}_{s}(z)}{\mu_{s}^{0}}=-\hat{y}\frac{\sinh\frac{2z-d_{N}%
}{2\lambda}}{\sinh\frac{d_{N}}{2\lambda}}+\left[  \hat{m}\times\left(  \hat
{m}\times\hat{y}\right)  \operatorname{Re}+\left(  \hat{m}\times\hat
{y}\right)  \operatorname{Im}\right]  \frac{2\lambda G_{\uparrow\downarrow}%
}{\sigma+2\lambda G_{\uparrow\downarrow}\coth\frac{d_{N}}{\lambda}}\frac
{\cosh\frac{z-d_{N}}{\lambda}}{\sinh\frac{d_{N}}{\lambda}}, \label{musgen}%
\end{equation}
then leads to the distributed spin current in N
\begin{equation}
\frac{\vec{j}_{s}^{z}(z)}{j_{s0}^{\mathrm{SH}}}=\hat{y}\frac{\cosh
\frac{2z-d_{N}}{2\lambda}-\cosh\frac{d_{N}}{2\lambda}}{\cosh\frac{d_{N}%
}{2\lambda}}-\left[  \hat{m}\times\left(  \hat{m}\times\hat{y}\right)
\operatorname{Re}+\left(  \hat{m}\times\hat{y}\right)  \operatorname{Im}%
\right]  \frac{2\lambda G_{\uparrow\downarrow}\tanh\frac{d_{N}}{2\lambda}%
}{\sigma+2\lambda G_{\uparrow\downarrow}\coth\frac{d_{N}}{\lambda}}\frac
{\sinh\frac{z-d_{N}}{\lambda}}{\sinh\frac{d_{N}}{\lambda}}.
\end{equation}
The ISHE drives a charge current in the $x$-$y$ plane by the diffusion spin
current component flowing along the $\hat{z}$-direction. The total
longitudinal (along $\hat{x}$) and transverse or Hall (along $\hat{y}$) charge
currents become
\begin{align}
\frac{j_{c,\mathrm{long}}(z)}{j_{c}^{0}}  &  =1+\theta_{\mathrm{SH}}%
^{2}\left[  \frac{\cosh\frac{2z-d_{N}}{2\lambda}}{\cosh\frac{d_{N}}{2\lambda}%
}+\left(  1-m_{y}^{2}\right)  \operatorname{Re}\frac{2\lambda G_{\uparrow
\downarrow}\tanh\frac{d_{N}}{2\lambda}}{\sigma+2\lambda G_{\uparrow\downarrow
}\coth\frac{d_{N}}{\lambda}}\frac{\sinh\frac{z-d_{N}}{\lambda}}{\sinh
\frac{d_{N}}{\lambda}}\right]  ,\\
\frac{j_{c,\mathrm{trans}}(z)}{j_{c}^{0}}  &  =\theta_{\mathrm{SH}}^{2}\left(
m_{x}m_{y}\operatorname{Re}-m_{z}\operatorname{Im}\right)  \frac{2\lambda
G_{\uparrow\downarrow}\tanh\frac{d_{N}}{2\lambda}}{\sigma+2\lambda
G_{\uparrow\downarrow}\coth\frac{d_{N}}{\lambda}}\frac{\sinh\frac{z-d_{N}%
}{\lambda}}{\sinh\frac{d_{N}}{\lambda}},
\end{align}
where $j_{c}^{0}=\sigma E_{x}$ is the charge current driven by the external
electric field.

The charge current vector is the observable in the experiment that is usually
expressed in terms of the longitudinal and transverse (Hall) resistivities.
Averaging the electric currents over the film thickness $z$ and expanding the
longitudinal resistivity governed by the current in the ($x$-)direction of the
applied field to leading order in $\theta_{\mathrm{SH}}^{2}$, we obtain
\begin{align}
\rho_{\mathrm{long}}  &  =\sigma_{\mathrm{long}}^{-1}=\left(  \frac
{\overline{j_{c,\mathrm{long}}}}{E_{x}}\right)  ^{-1}\approx\rho+\Delta
\rho_{0}+\Delta\rho_{1}\left(  1-m_{y}^{2}\right)  ,\\
\rho_{\mathrm{trans}}  &  =-\frac{\sigma_{\mathrm{trans}}}{\sigma
_{\mathrm{long}}^{2}}\approx-\frac{\overline{j_{c,\mathrm{trans}}}/E_{x}%
}{\sigma^{2}}=\Delta\rho_{1}m_{x}m_{y}+\Delta\rho_{2}m_{z},
\end{align}
where
\begin{align}
\frac{\Delta\rho_{0}}{\rho}  &  =-\theta_{\mathrm{SH}}^{2}\frac{2\lambda
}{d_{N}}\tanh\frac{d_{N}}{2\lambda},\label{rho-0}\\
\frac{\Delta\rho_{1}}{\rho}  &  =\theta_{\mathrm{SH}}^{2}\frac{\lambda}{d_{N}%
}\operatorname{Re}\frac{2\lambda G_{\uparrow\downarrow}\tanh^{2}\frac{d_{N}%
}{2\lambda}}{\sigma+2\lambda G_{\uparrow\downarrow}\coth\frac{d_{N}}{\lambda}%
},\\
\frac{\Delta\rho_{2}}{\rho}  &  =-\theta_{\mathrm{SH}}^{2}\frac{\lambda}%
{d_{N}}\operatorname{Im}\frac{2\lambda G_{\uparrow\downarrow}\tanh^{2}%
\frac{d_{N}}{2\lambda}}{\sigma+2\lambda G_{\uparrow\downarrow}\coth\frac
{d_{N}}{\lambda}},
\end{align}
where $\rho=\sigma^{-1}$ is the intrinsic electric resistivity of the bulk
normal metal. $\Delta\rho_{0}<0$ seems to imply that the resistivity is
reduced by the spin-orbit interaction. However, this is an effect of the order
of $\theta_{\mathrm{SH}}^{2}$ that becomes relevant only when $d_{N}\ $is
sufficiently small. The spin-orbit interaction also generates spin-flip
scattering that increases the resistance to leading order according to
Matthiesen's rule. We see that $\Delta\rho_{1}$ (caused mainly by $G_{r}$)
contributes to the SMR, while $\Delta\rho_{2}$ (caused mainly by $G_{i}$)
contributes only when there is a magnetization component normal to the plane
(AHE), as discussed below.

\subsection{Limit of $G_{i}=\operatorname{Im}G_{\uparrow\downarrow}%
\ll\operatorname{Re}G_{\uparrow\downarrow}=G_{r}$}

\begin{figure}[ptb]
\includegraphics[width=0.48\textwidth,angle=0]{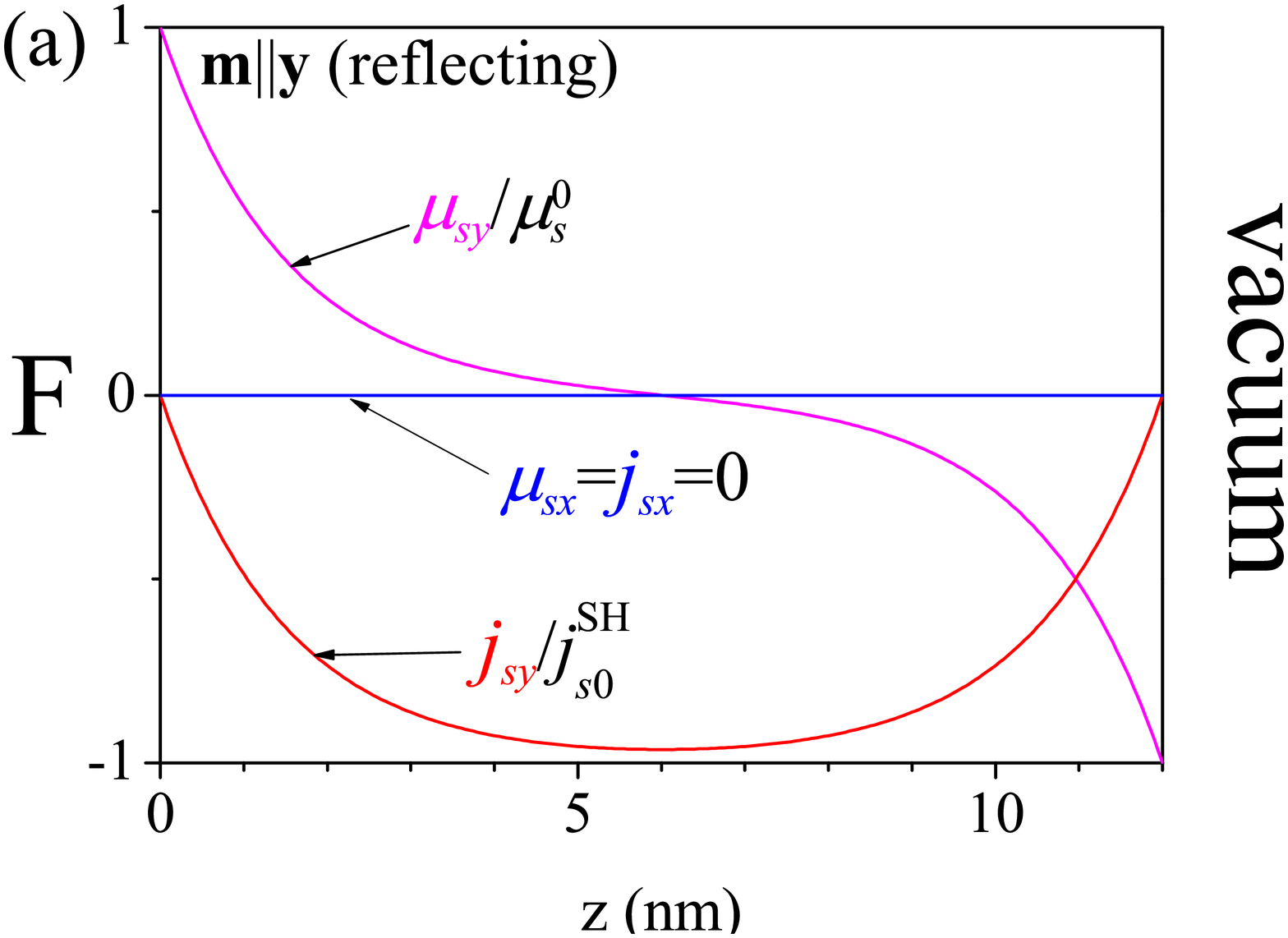}
\includegraphics[width=0.48\textwidth,angle=0]{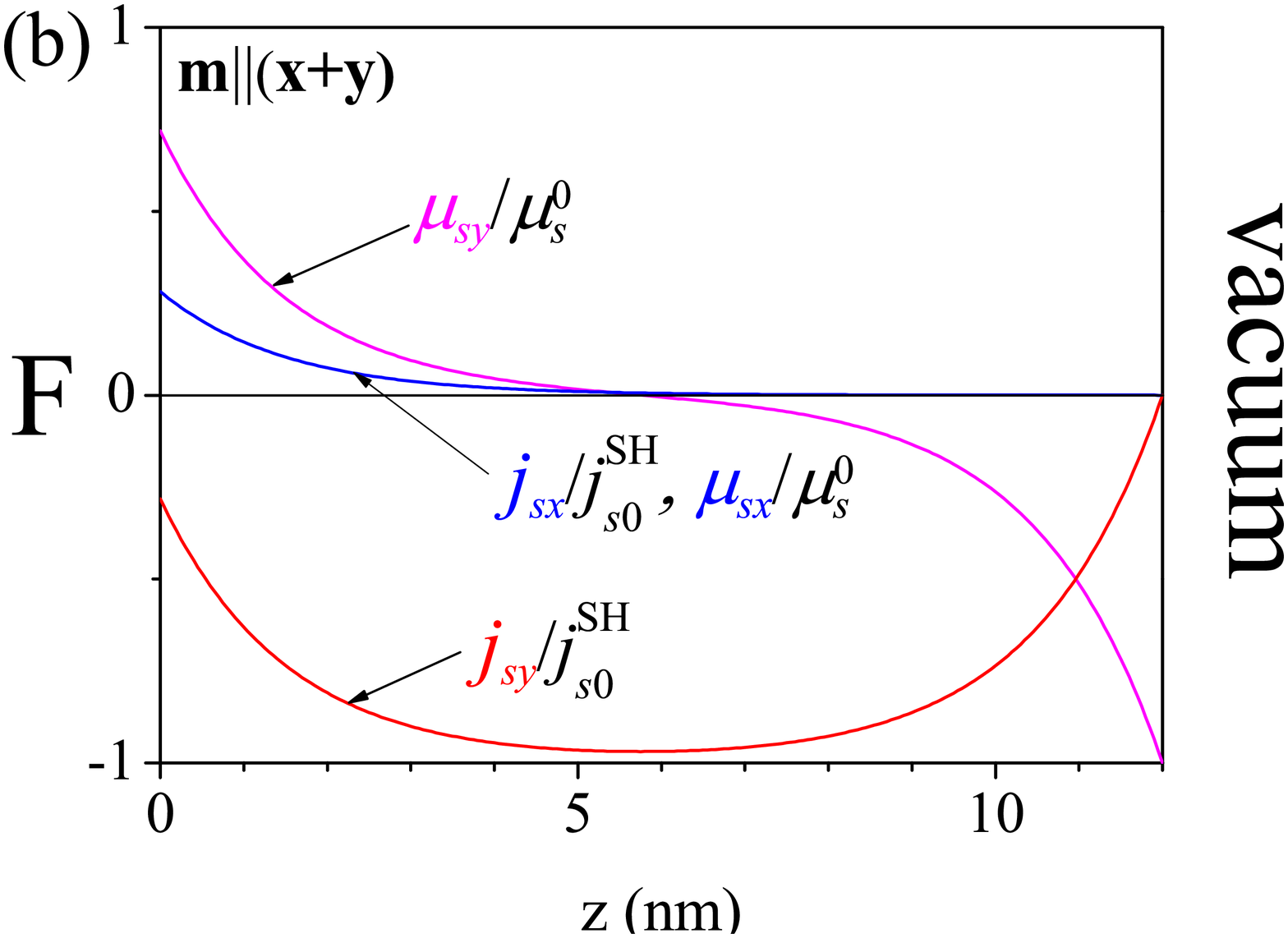}
\includegraphics[width=0.48\textwidth,angle=0]{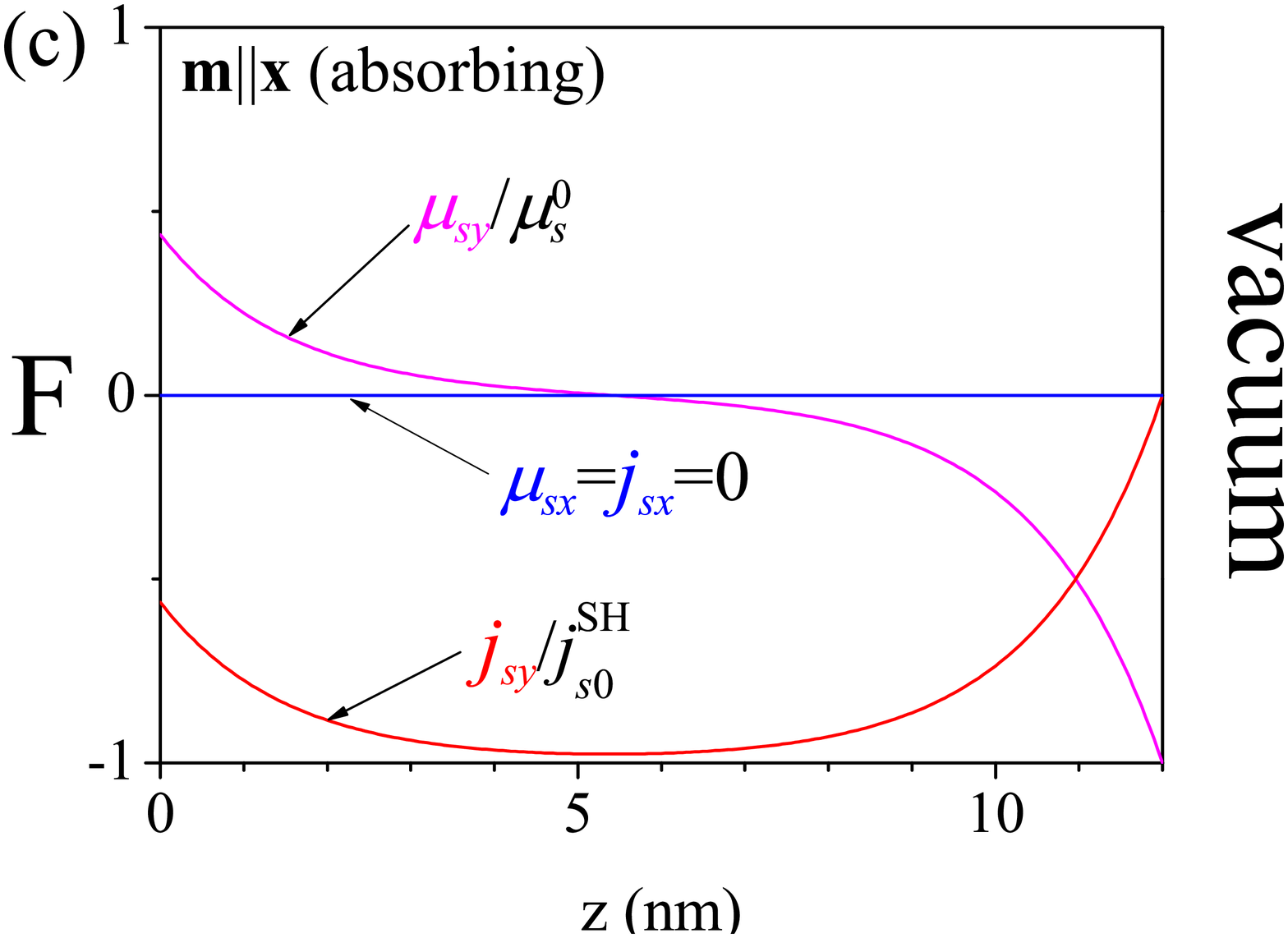}
\caption{(Color online). Normalized $\mu_{sx}$, $\mu_{sy}$, $j_{sx}$, and
$j_{sy}$ as functions of $z$ for magnetizations (a) $\hat{m}=\hat{y}$, (b)
$\hat{m}=\left(  \hat{x}+\hat{y}\right)  /\sqrt{2}$, and (c) $\hat{m}=\hat{x}$
for a sample with $d_{N}=12$~nm. We adopt the transport parameters
$\rho=8.6\times10^{-7}\operatorname{\Omega}\operatorname{m}$, $\lambda
=1.5$~nm, and $G_{r}=5\times10^{14}\operatorname{\Omega}^{-1}%
\operatorname{m}^{-2}$. For magnetizations $\hat{m}=\hat{y}$ and $\hat{m}%
=\hat{x}$, both $\mu_{sx}$ and $j_{sx}$ are $0$.}%
\label{jsx_jsy_z-dep_1}%
\end{figure}
According to first principles calculations,\cite{Jia11} $\left\vert
G_{i}\right\vert $ is at least one order of magnitude smaller than $G_{r}$ for
YIG, so $G_{i}=0$ appears to be a good first approximation. In this limit, we
plot normalized components of spin accumulation ($\mu_{sx}$ and $\mu_{sy}$)
and spin current ($j_{sx}=\vec{j}_{s}^{z}\cdot\hat{x}$ and $j_{sy}=\vec{j}%
_{s}^{z}\cdot\hat{y}$) as functions of $z$ for different magnetizations in
Fig~\ref{jsx_jsy_z-dep_1}. When the magnetization of F is along $\hat{y}$, the
spin current at the N$|$F interface ($z=0$) vanishes just as for the vacuum
interface. By rotating the magnetization from $\hat{y}$ to $\hat{x}$, the spin
current at the N$|$F interface and the torque on the magnetization is
activated, while the spin accumulation is dissipated correspondingly. We note
that the $x$-components of both spin accumulation and spin current vanish when
the magnetization is along $\hat{x}$ and $\hat{y}$, and reach a maximum value
at $\left(  \hat{x}+\hat{y}\right)  /\sqrt{2}$.

For $G_{i}=0$ the observable transport properties reduce to
\begin{align}
\rho_{\mathrm{long}}  &  \approx\rho+\Delta\rho_{0}+\Delta\rho_{1}\left(
1-m_{y}^{2}\right)  ,\label{rho_long_1}\\
\rho_{\mathrm{trans}}  &  \approx\Delta\rho_{1}m_{x}m_{y}, \label{rho_trans_1}%
\end{align}
where
\begin{align}
\frac{\Delta\rho_{0}}{\rho}  &  =-\theta_{\mathrm{SH}}^{2}\frac{2\lambda
}{d_{N}}\tanh\frac{d_{N}}{2\lambda},\\
\frac{\Delta\rho_{1}}{\rho}  &  =\theta_{\mathrm{SH}}^{2}\frac{\lambda}{d_{N}%
}\frac{2\lambda G_{r}\tanh^{2}\frac{d_{N}}{2\lambda}}{\sigma+2\lambda
G_{r}\coth\frac{d_{N}}{\lambda}}. \label{Drho_1}%
\end{align}
Equations~(\ref{rho_long_1}-\ref{rho_trans_1}) fully explain the magnetization
dependence of SMR in Ref.~\onlinecite{Nakayama12}, while Eq.~(\ref{Drho_1})
shows that an SMR exists only when the spin-mixing conductance does not
vanish. Since results do not depend on the $z$-component of magnetization, the
AHE vanishes in our model when $G_{i}=0$.

\subsection{$G_{r}\gg\sigma/\left(  2\lambda\right)  $}

Here we discuss the limit in which the spin current transverse to $\hat{m}$ is
completely absorbed as an STT without reflection. This ideal situation is
actually not so far from reality for the recently found large $G_{r}$ between
YIG\ and noble metals.\cite{Jia11,Heinrich11} The spin current at the
interface is then
\begin{equation}
\frac{\vec{j}_{s}^{(\mathrm{F})}}{j_{s0}^{\mathrm{SH}}}\overset{G_{r}\gg
\sigma/\left(  2\lambda\right)  }{=}\hat{m}\times\left(  \hat{m}\times\hat
{y}\right)  \tanh\frac{d_{N}}{\lambda}\tanh\frac{d_{N}}{2\lambda},
\end{equation}
and the maximum magnetoresistance for the bilayer is
\begin{equation}
\frac{\Delta\rho_{1}}{\rho}=\theta_{\mathrm{SH}}^{2}\frac{\lambda}{d_{N}}%
\tanh\frac{d_{N}}{\lambda}\tanh^{2}\frac{d_{N}}{2\lambda}.
\end{equation}
In Sec.~\ref{comparison} we test this limit with available parameters from experiments.

\subsection{$\lambda/d_{N}\gg1$}

When the spin-diffusion length is much larger than the thickness of N%
\begin{equation}
\frac{\vec{\mu}_{s}(z)}{\mu_{s}^{0}}\overset{\lambda/d_{N}\gg1}{=}\hat
{m}\times\left(  \hat{m}\times\hat{y}\right)  -\hat{y}\frac{2z-d_{N}}{d_{N}%
},\nonumber
\end{equation}
while spin current and magnetoresistance vanish. We can interpret this as
multiple scattering of the spin current at the interfaces; the ISHE has both
positive and negative charge current contributions that cancel each other.

\subsection{Spin Hall AHE}

Recent measurements in YIG$|$Pt display a small AHE-like signal on top of the
ordinary Hall effect, \textit{i.e}. a transverse voltage when the
magnetization is normal to the film.\cite{Althammer12} As mentioned above, an
imaginary part of the spin-mixing conductance $G_{i}$ can cause a spin Hall
AHE (SHAHE).

The component of the spin accumulation $\mu_{sx}$
\begin{equation}
\frac{\mu_{sx}(z)}{\mu_{s}^{0}}=\frac{2\lambda}{\sigma}\frac{\cosh
\frac{z-d_{N}}{\lambda}}{\sinh\frac{d_{N}}{\lambda}}\left[  m_{x}%
m_{y}\operatorname{Re}-m_{z}\operatorname{Im}\right]  \frac{\sigma G_{\uparrow
\downarrow}}{\sigma+2\lambda G_{\uparrow\downarrow}\coth\frac{d_{N}%
}{\lambda}}%
\end{equation}
contains a contribution that scales with $m_{z}$ and contributes a charge
current in the transverse ($\hat{y}$-) direction
\begin{equation}
\frac{j_{c,\mathrm{trans}}^{(\mathrm{SHAHE})}(z)}{j_{c}^{0}}=-2\lambda
\theta_{\mathrm{SH}}^{2}m_{z}\frac{\sinh\frac{z-d_{N}}{\lambda}}{\sinh
\frac{d_{N}}{\lambda}}\operatorname{Im}\frac{G_{\uparrow\downarrow}\tanh
\frac{d_{N}}{2\lambda}}{\sigma+2\lambda G_{\uparrow\downarrow}\coth\frac
{d_{N}}{\lambda}}.
\end{equation}
The transverse resistivity due to this current is
\begin{equation}
\rho_{\mathrm{trans}}^{(\mathrm{SHAHE})}\approx-\frac{\overline
{j_{c,\mathrm{trans}}^{(\mathrm{SHAHE})}}/E_{x}}{\sigma^{2}}=-\Delta\rho
_{2}m_{z},
\end{equation}
where
\begin{equation}
\frac{\Delta\rho_{2}}{\rho}\approx\frac{2\lambda^{2}\theta_{\mathrm{SH}}^{2}%
}{d_{N}}\frac{\sigma G_{i}\tanh^{2}\frac{d_{N}}{2\lambda}}{\left(
\sigma+2\lambda G_{r}\coth\frac{d_{N}}{\lambda}\right)  ^{2}+\left(  2\lambda
G_{i}\coth\frac{d_{N}}{\lambda}\right)  ^{2}}\approx\frac{2\lambda^{2}%
\theta_{\mathrm{SH}}^{2}}{d_{N}}\frac{\sigma G_{i}\tanh^{2}\frac{d_{N}%
}{2\lambda}}{\left(  \sigma+2\lambda G_{r}\coth\frac{d_{N}}{\lambda}\right)
^{2}}.\nonumber
\end{equation}

\subsection{Comparison with experiments}

\label{comparison}

\begin{figure}[ptb]
\includegraphics[width=0.45\textwidth,angle=0]{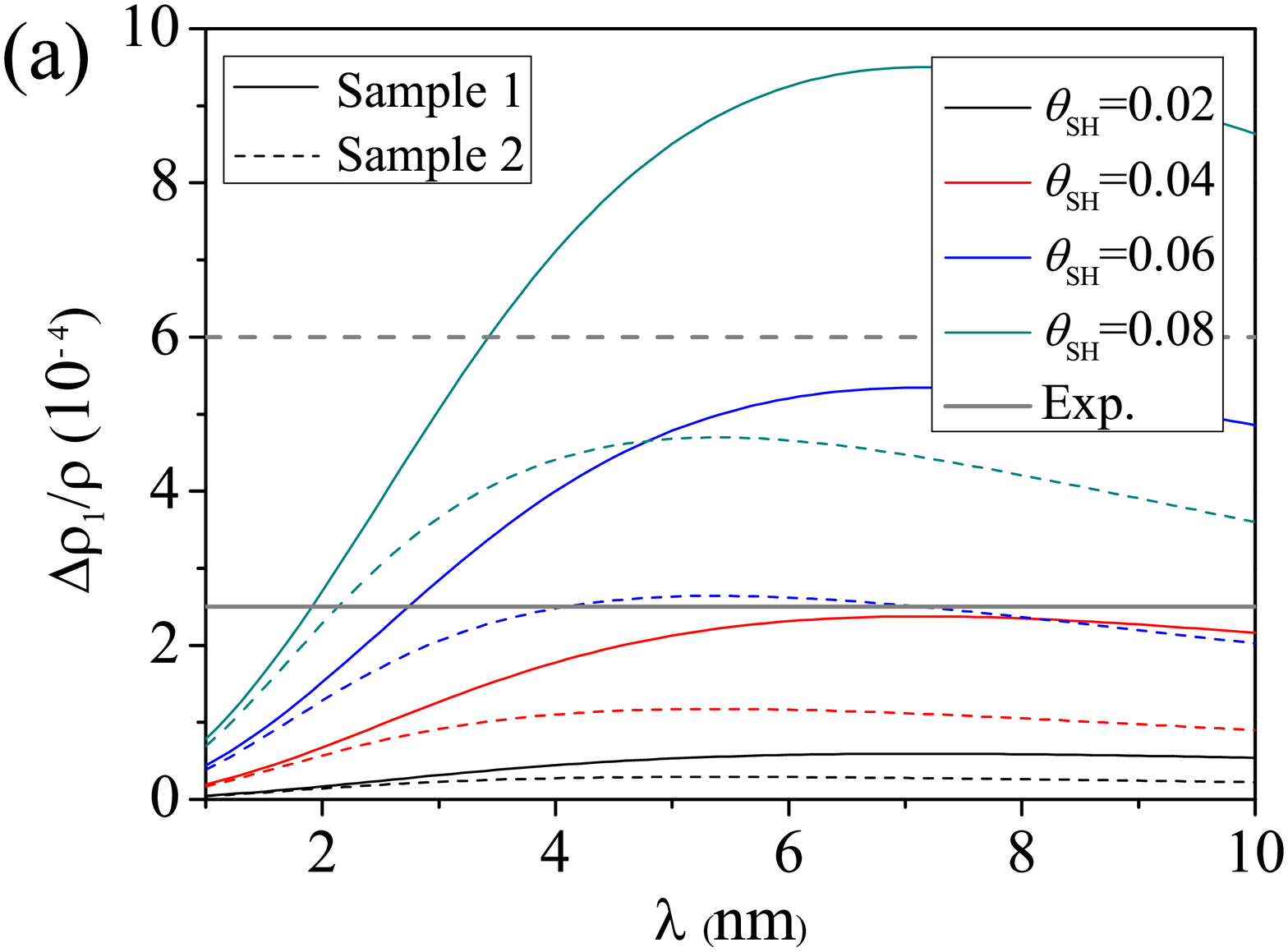}
\includegraphics[width=0.45\textwidth,angle=0]{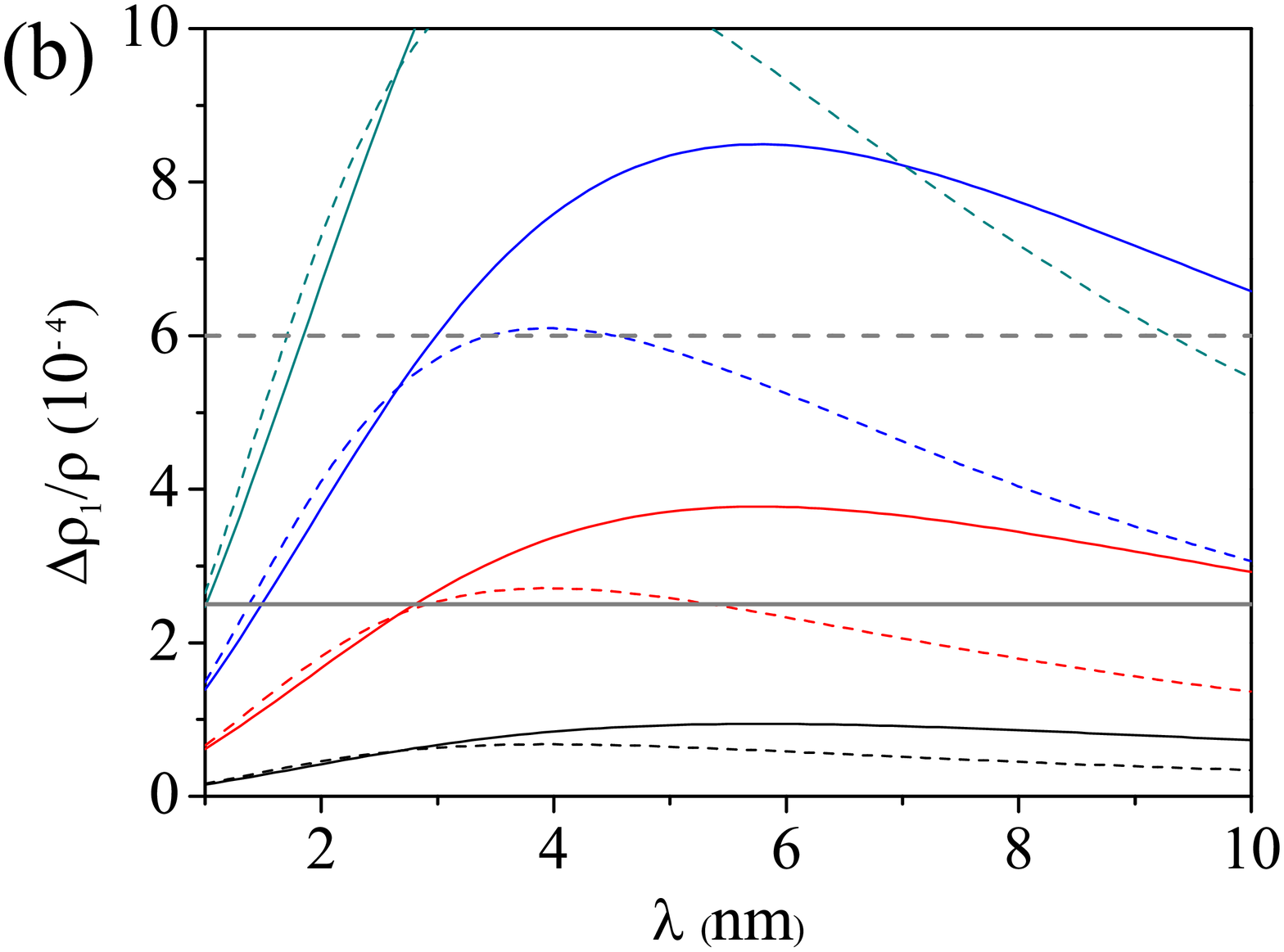}
\includegraphics[width=0.45\textwidth,angle=0]{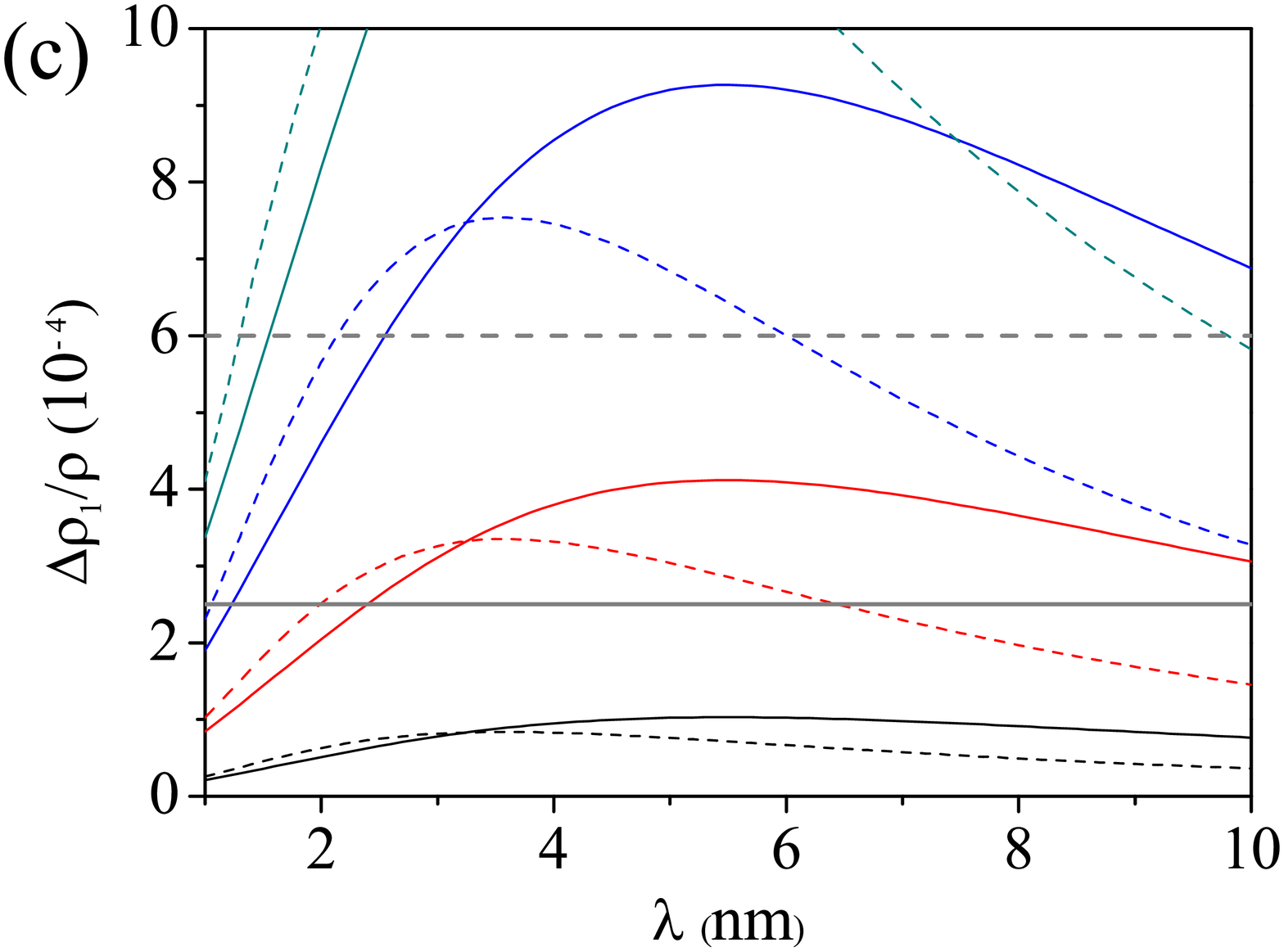}
\includegraphics[width=0.45\textwidth,angle=0]{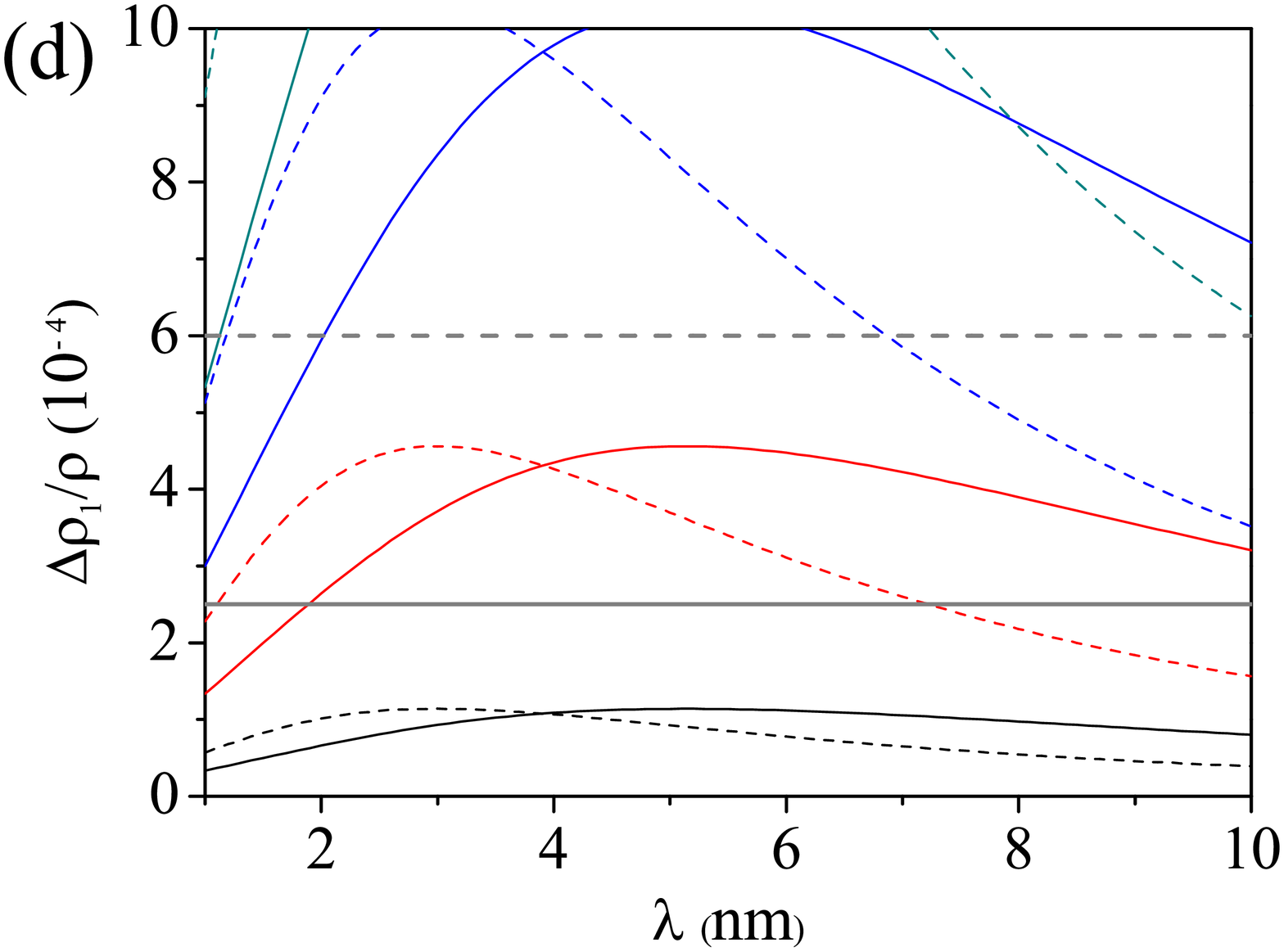}
\caption{(Color online) Calculated $\Delta\rho_{1}/\rho$ as a function of
$\lambda$ for different spin Hall angles $\theta_{\mathrm{SH}}$ with
(a)~$G_{r}=1\times10^{14}\operatorname{\Omega}^{-1}\operatorname{m}^{-2}$,
(b)~$G_{r}=5\times10^{14}\operatorname{\Omega}^{-1}\operatorname{m}^{-2}$,
(c)~$G_{r}=10\times10^{14}\operatorname{\Omega}^{-1}\operatorname{m}^{-2}$,
and (d) the ideal limit $G_{r}\gg\sigma/(2\lambda)$. The Pt layers are
12-nm-thick with resistivity $8.6\times10^{-7}\operatorname{\Omega
}\operatorname{m}$ (Sample 1, solid curve) and 7-nm-thick with resistivity
$4.1\times10^{-7}\operatorname{\Omega}\operatorname{m}$ (Sample 2, dashed
curve). Experimental results are shown as horizontal lines for
comparison.\cite{Nakayama12}}%
\label{SMR-different-parameters}%
\end{figure}

There are controversies about the values of the material parameters relevant
for our theory, \textit{i.e}. the spin-mixing conductance $G_{\uparrow
\downarrow}$ of the N$|$F interface, as well as spin-flip diffusion length
$\lambda$ and spin Hall angle $\theta_{\mathrm{SH}}$ in the normal metal.

Experimentally, Burrows \textit{et al.}\cite{Heinrich11} found for an Au$|$YIG
interface with $G_{0}=e^{2}/h.$%
\begin{equation}
\frac{G_{r}^{\mathrm{exp}}}{G_{0}}=5.2\times10^{18}\operatorname{m}%
^{-2};\;G_{r}^{\mathrm{exp}}=2\times10^{14}\mathrm{\Omega}^{-1}%
\operatorname{m}^{-2}.
\end{equation}
On the theory side, the spin-mixing conductance from scattering theory for an
insulator reads\cite{Brataas00}
\begin{equation}
\frac{G_{\uparrow\downarrow}}{G_{0}}=N_{\mathrm{Sh}}-\sum_{n}r_{n\uparrow
}^{\ast}r_{n\downarrow}=N_{\mathrm{Sh}}-\sum_{n}e^{i\left(  \delta
_{n\downarrow}-\delta_{n\uparrow}\right)  }, \label{g-mix}%
\end{equation}
where $r_{n\uparrow(\downarrow)}=e^{i\delta_{n\uparrow(\downarrow)}}$ is the
reflection coefficient of an electron in the quantum channel $n$ on a unit
area at the N$|$F interface with unit modulus and phase $\delta_{n\uparrow
(\downarrow)}$ for the majority (minority) spin, and $N_{\mathrm{Sh}}$ is the
number of transport channels (per unit area) at the Fermi energy,
\textit{i.e.} $N_{\mathrm{Sh}}$ is the Sharvin conductance (for one spin).
Therefore
\begin{equation}
\frac{G_{r}}{G_{0}}\leq2N_{\mathrm{Sh}};\;\frac{\left\vert G_{i}\right\vert
}{G_{0}}\leq N_{\mathrm{Sh}},
\end{equation}

Jia \textit{et al}.\cite{Jia11} computed Eq.~(\ref{g-mix}) for a Ag$|$YIG
interface by first principles. The average of different crystal interfaces
\begin{equation}
G_{r}^{(0)}=2.3\times10^{14}\operatorname{\Omega}^{-1}\operatorname{m}^{-2},
\end{equation}
is quite close to the Sharvin conductance of silver ($N_{\mathrm{Sh}}%
G_{0}\approx4.5\times10^{14}\mathrm{\Omega}^{-1}\operatorname{m}^{-2}$).

For comparison with experiment we have to include the Schep drift
correction:\cite{Schep97}
\begin{equation}
\frac{1}{\tilde{G}_{r}/G_{0}}=\frac{1}{G_{r}^{(0)}/G_{0}}-\frac{1}%
{2N_{\mathrm{Sh}}},
\end{equation}
which leads to
\begin{equation}
\tilde{G}_{r}\approx3.1 \times10^{14}\operatorname{\Omega}^{-1}%
\operatorname{m}^{-2}.
\end{equation}
One should note that the mixing conductance of the Pt$|$YIG interface can then
be estimated to be $\tilde{G}_{r}\approx10^{15}\operatorname{\Omega}
^{-1}\operatorname{m}^{-2}$ since the Pt conduction electron density and
Sharvin conductance are higher than those of noble metals.

Using parameters $\rho=\sigma^{-1}=8.6\times10^{-7}\operatorname{\Omega
}\operatorname{m}$, $d_{N}=12$~nm, and $\lambda=1.5$~nm,\cite{Liu11} we see
that the absorbed transverse spin currents with $G_{r}=\tilde{G}_{r}$ and
$G_{r}=G_{r}^{\mathrm{max}}$ obtained from above for a Ag$|$YIG interface are
$44\%$ and $70\%$ of the value for a perfect spin sink $G_{r}\rightarrow
\infty$, respectively. For a Pt$|$YIG interface this value should be even larger.

In order to compare our results with the observed SMR, we have to fill in or
fit the parameters. The values of the spin-diffusion length and the spin Hall angle
differ widely.\cite{Liu11} In Fig.~\ref{SMR-different-parameters} we plot the
SMR for three fixed values of $G_{r}$. We observe that the experiments can be
explained by a sensible set of transport parameters ($G_{r}$, $\lambda$,
$\theta_{\mathrm{SH}}$) that somewhat differ for the two representative
samples reported in Ref.~\onlinecite{Nakayama12}. Generally, the SMR increases
with a larger value of $G_{r}$ but decreases when $\lambda$ is getting longer.
These features are in agreement with the discussion of the simple limits
above. Sample 1 in Ref.~\onlinecite{Nakayama12} has a larger resistivity but a
smaller SMR (ratio), implying a smaller spin Hall angle and/or smaller
spin-diffusion length. When we fix the spin Hall angle $\theta_{\mathrm{SH}}=0.06$
and the spin-mixing conductance $G_{r}=5\times10^{14}\operatorname{\Omega }^{-1}
\operatorname{m}^{-2}$, the corresponding estimated spin-diffusion lengths of Samples 1 and 2 are
$\lambda_{1}\approx1.5\,$nm and $\lambda_{2}\approx3.5\,$nm, respectively.

Finally we discuss the AHE equivalent or SHAHE. From experiments $\Delta
\rho_{2}/\rho\approx1.5\times10^{-5}$ for $\rho=4.1\times10^{-7}%
\mathrm{\Omega}\operatorname{m}$ and $d_{N}=7\,\operatorname{nm}%
$.\cite{Althammer12} Choosing $\theta_{\mathrm{SH}}=0.05$, $\lambda=1.5\,$nm,
and $G_{r}=5\times10^{14}\mathrm{\Omega}^{-1}\operatorname{m}^{-2}$, we would
need a $G_{i}=6.2\times10^{13}\mathrm{\Omega}^{-1}\operatorname{m}^{-2}$ to
explain experiments, a number that is supported by first principle
calculations.\cite{Jia11}

\section{Spin valves}

\label{yig-pt-yig} In this section we discuss F$\left(  \hat{m}\right)  |$%
N$|$F$\left(  \hat{m}^{\prime}\right)  $ spin valves fabricated from magnetic
insulators with magnetization directions $\hat{m}$ and $\hat{m}^{\prime}$. The
general angle dependence for independent rotations of $\hat{m}$ and $\hat
{m}^{\prime}$ is straightforward but tedious. We discuss in the following two
representative configurations in which the two magnetizations are parallel and
perpendicular to each other. We disregard in the following the effective field
due to $G_{i}$ such that the parallel and antiparallel configurations $\hat
{m}=\pm\hat{m}^{\prime}$ are equivalent. Moreover, we limit the discussion to
the simple case of two identical F$|$N and N$|$F interfaces, \textit{i.e.},
the spin-mixing conductances at both interfaces are the same.

\subsection{Parallel Configuration ($\hat{m}\cdot\hat{m}^{\prime}=\pm1$)}

When the magnetizations are aligned in parallel or antiparallel configuration,
the boundary condition $\vec{j}_{s}^{(z)}(d_{N})=-\vec{j}_{s}^{(\mathrm{F})}$
applies. We proceed as in Sec.~\ref{vac-pt-yig} to obtain the spin
accumulation
\begin{equation}
\frac{\vec{\mu}_{s}}{\mu_{s}^{0}}=-\left[  \hat{y}+\hat{m}\times\left(
\hat{m}\times\hat{y}\right)  \frac{2\lambda G_{r}\tanh\frac{d_{N}}{2\lambda}%
}{\sigma+2\lambda G_{r}\tanh\frac{d_{N}}{2\lambda}}\right]  \frac{\sinh
\frac{2z-d_{N}}{2\lambda}}{\sinh\frac{d_{N}}{2\lambda}},
\end{equation}
and the spin current
\begin{equation}
\frac{\vec{j}_{s}^{z}}{j_{s0}^{\mathrm{SH}}}=\hat{y}\left(  \frac{\cosh
\frac{2z-d_{N}}{2\lambda}}{\cosh\frac{d_{N}}{2\lambda}}-1\right)  +\hat
{m}\times\left(  \hat{m}\times\hat{y}\right)  \frac{2\lambda G_{r}\tanh
\frac{d_{N}}{2\lambda}}{\sigma+2\lambda G_{r}\tanh\frac{d_{N}}{2\lambda}}%
\frac{\cosh\frac{2z-d_{N}}{2\lambda}}{\cosh\frac{d_{N}}{2\lambda}}.\nonumber
\end{equation}
The spin currents at the bottom and top of N are absorbed as STTs and
read
\begin{equation}
\frac{\vec{j}_{s}^{z}(0)}{j_{s0}^{\mathrm{SH}}}=\frac{\vec{j}_{s}^{z}(d_{N}%
)}{j_{s0}^{\mathrm{SH}}}=\hat{m}\times\left(  \hat{m}\times\hat{y}\right)
\frac{2\lambda G_{r}\tanh\frac{d_{N}}{2\lambda}}{\sigma+2\lambda G_{r}%
\tanh\frac{d_{N}}{2\lambda}},
\end{equation}
leading to opposite STTs at the bottom ($\vec{\tau}_{\mathrm{stt}%
}^{\mathrm{(B)}}$) and top ($\vec{\tau}_{\mathrm{stt}}^{\mathrm{(T)}}$)
ferromagnets
\begin{equation}
\vec{\tau}_{\mathrm{stt}}^{\mathrm{(B)}}=\frac{\hbar}{2e}\vec{j}_{s}%
^{(z)}(0)=-\vec{\tau}_{\mathrm{stt}}^{\mathrm{(T)}}%
\end{equation}
since $\vec{j}_{s}^{(\mathrm{F})}(\hat{m})=\vec{j}_{s}^{z}(0)=\vec{j}_{s}%
^{z}(d_{N})=-\vec{j}_{s}^{(\mathrm{F})}(\hat{m}^{\prime})$.

\begin{figure}[ptb]
\includegraphics[width=0.45\textwidth,angle=0]{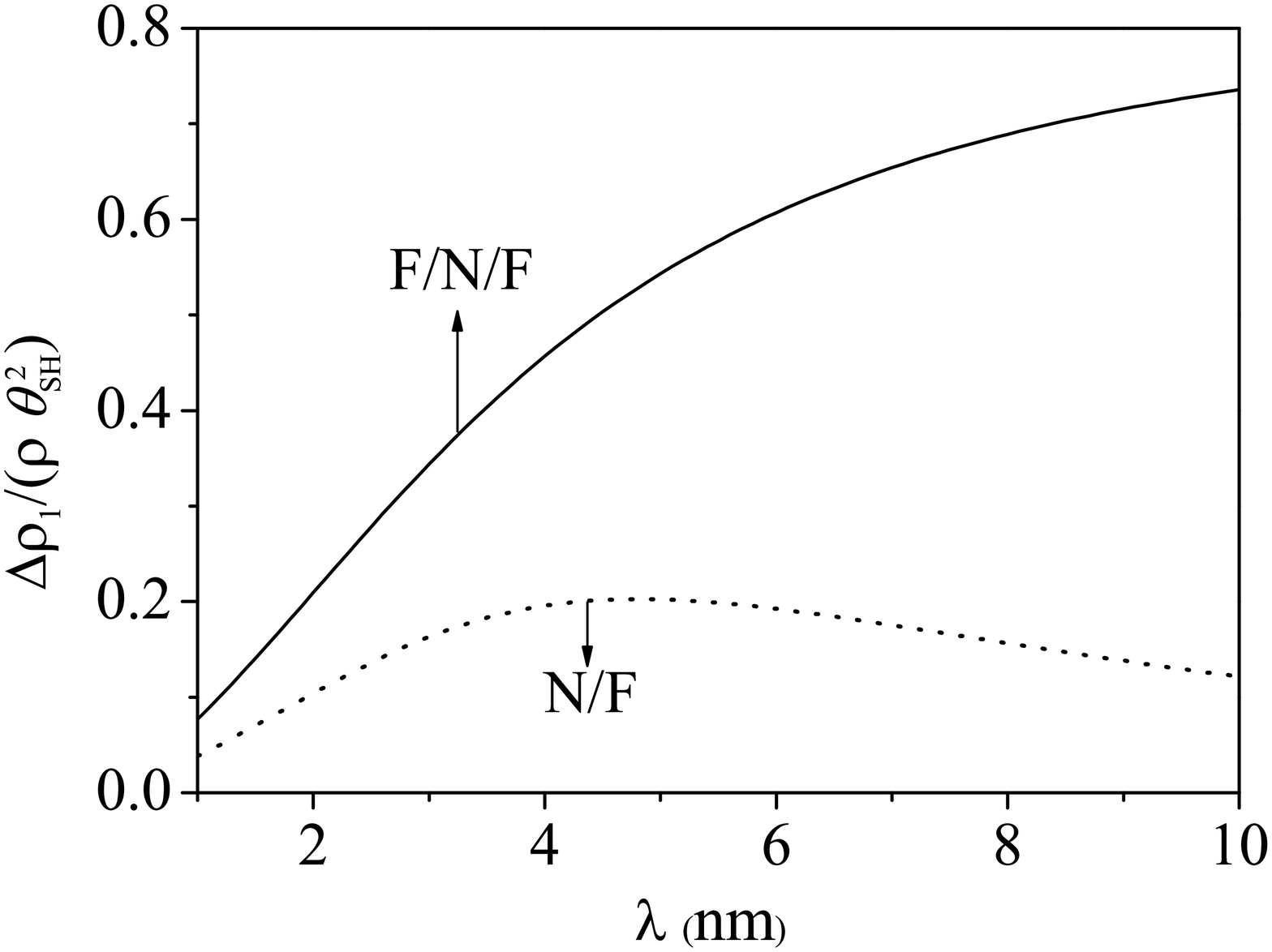}
\caption{(Color online) Calculated $\Delta\rho_{1}/\left(  \rho\theta_{\mathrm{SH}}^{2}\right)$ in an F$|$N$|$F spin valve as a function of spin-diffusion length $\lambda$ with $d_{N}=12$~nm, $G_{r}%
=5\times10^{14}\operatorname{\Omega}^{-1}\mathrm{\operatorname{m}}^{-2}$, and
$\rho=8.6\times10^{-7}\operatorname{\Omega}\mathrm{\operatorname{m}}$
chosen from Sample 1 in Ref.~\onlinecite{Nakayama12}. $\Delta\rho_{1}/\left(  \rho\theta_{\mathrm{SH}}^{2}\right)$ in
an N$\vert$F bilayer is plotted as a dotted line for comparison.}%
\label{SMR-trilayer}%
\end{figure}

The longitudinal and transverse (Hall) charge currents are
\begin{align}
\frac{j_{c,\mathrm{long}}}{j_{c}^{0}}  &  =1+\theta_{\mathrm{SH}}^{2}\left[
1-\left(  1-m_{y}^{2}\right)  \frac{2\lambda G_{r}\tanh\frac{d_{N}}{2\lambda}%
}{\sigma+2\lambda G_{r}\tanh\frac{d_{N}}{2\lambda}}\right]  \frac{\cosh
\frac{2z-d_{N}}{2\lambda}}{\cosh\frac{d_{N}}{2\lambda}},\\
\frac{j_{c,\mathrm{trans}}}{j_{c}^{0}}  &  =-\theta_{\mathrm{SH}}^{2}%
m_{x}m_{y}\frac{2\lambda G_{r}\tanh\frac{d_{N}}{2\lambda}}{\sigma+2\lambda
G_{r}\tanh\frac{d_{N}}{2\lambda}}\frac{\cosh\frac{2z-d_{N}}{2\lambda}}%
{\cosh\frac{d_{N}}{2\lambda}}.
\end{align}
and the longitudinal and transverse resistivities read
\begin{align}
\rho_{\mathrm{long}}  &  =\rho+\Delta\rho_{0}+\Delta\rho_{1}\left(
1-m_{y}^{2}\right)  ,\\
\rho_{\mathrm{trans}}  &  =\Delta\rho_{1} m_{x}m_{y},
\end{align}
where
\begin{align}
\frac{\Delta\rho_{0}}{\rho}  &  =-\theta_{\mathrm{SH}}^{2}\frac{2\lambda
}{d_{N}}\tanh\frac{d_{N}}{2\lambda},\\
\frac{\Delta\rho_{1}}{\rho}  &  =\frac{\theta_{\mathrm{SH}}^{2}}{d_{N}}%
\frac{4\lambda^{2}G_{r}\tanh^{2}\frac{d_{N}}{2\lambda}}{\sigma+2\lambda
G_{r}\tanh\frac{d_{N}}{2\lambda}}.
\end{align}
Figure \ref{SMR-trilayer} shows $\Delta\rho_{1}/\left(  \rho\theta_{\mathrm{SH}}^{2}\right)$ with respect to the
spin-diffusion length in an F$|$N$|$F spin valve
with parallel magnetization configuration. Compared to N$|$F bilayers, the SMR
in spin valves is larger and does not vanish in the limit of long spin-diffusion lengths.

\subsection{Limit $\lambda/d_{N}\gg1$}

The spin accumulation for weak spin-flip reads
\begin{equation}
\frac{\vec{\mu}_{s}}{\mu_{s}^{0}}\overset{\lambda/d_{N}\gg1}{=}-\left[
\hat{y}+\frac{d_{N}G_{r}}{\sigma+d_{N}G_{r}}\hat{m}\times\left(  \hat{m}%
\times\hat{y}\right)  \right]  \frac{2z-d_{N}}{d_{N}},
\end{equation}
leading to the spin current
\begin{equation}
\frac{\vec{j}_{s}^{z}}{j_{s0}^{\mathrm{SH}}}\overset{\lambda/d_{N}\gg
1}{=}\frac{d_{N}G_{r}}{\sigma+d_{N}G_{r}}\hat{m}\times\left(  \hat{m}%
\times\hat{y}\right)  .
\end{equation}
In contrast to the bilayer, we find a finite SMR in this limit for spin
valves:%
\begin{align}
&  \frac{j_{c,\mathrm{long}}}{j_{c}^{0}}\overset{\lambda/d_{N}\gg1}{=}%
1+\theta_{\mathrm{SH}}^{2}\left[  1-\frac{d_{N}G_{r}}{\sigma+d_{N}G_{r}%
}\left(  1-m_{y}^{2}\right)  \right]  \overset{G_{r}\gg\sigma/d_{N}%
}{=}1+\theta_{\mathrm{SH}}^{2}m_{y}^{2},\\
&  \frac{j_{c,\mathrm{trans}}}{j_{c}^{0}}\overset{\lambda/d_{N}\gg1}{=}%
-\theta_{\mathrm{SH}}^{2}\frac{d_{N}G_{r}}{\sigma+d_{N}G_{r}}m_{x}%
m_{y}\overset{G_{r}\gg\sigma/d_{N}}{=}-\theta_{\mathrm{SH}}^{2}m_{x}m_{y}%
\end{align}
or%
\begin{align}
\frac{\Delta\rho_{0}}{\rho} &  =-\theta_{\mathrm{SH}}^{2},\\
\frac{\Delta\rho_{1}}{\rho} &  =\theta_{\mathrm{SH}}^{2}\frac{d_{N}G_{r}%
}{\sigma+d_{N}G_{r}}\overset{G_{r}\gg\sigma/d_{N}}{=}\theta_{\mathrm{SH}}^{2}.
\end{align}
Here we find the maximum achievable SMR effects in metals with spin Hall angle
$\theta_{\mathrm{SH}}$ by taking the limit of perfect spin current absorption.
Clearly this requires spin valves with sufficiently thin spacer layers. We
interpret these results in terms of spin angular momentum conservation: The
finite SMR\ is achieved by using the ferromagnet as a spin sink that
suppresses the back flow of spins and the ISHE. This process requires a source
of angular momentum, which in bilayers can only be the lattice of the normal
metal. Consequently, the SMR is suppressed in the F$|$N system when spin-flip
is not allowed. In spin valves, however, the second ferromagnet layer can act
as a spin current source, thereby allowing a finite SMR even in the absence of
spin-flip scattering.

\subsection{Perpendicular Configuration ($\hat{m}\cdot\hat{m}^{\prime}=0$)}

We may consider two in-plane magnetizations $\hat{m}=\left(  \cos\alpha
,\sin\alpha,0\right)  $ and $\hat{m}^{\prime}=\left(  -\sin\alpha,\cos
\alpha,0\right)  $, which are perpendicular to each other. When $\alpha=0$,
the first layer maximally absorbs the SHE spin current, while $\hat{m}%
^{\prime}$ is completely reflecting, just as the vacuum interface in the
bilayer. For general $\alpha$:
\begin{align}
\frac{\mu_{sx}(z)}{\mu_{s}^{0}} &  =\frac{2\lambda G_{r}}{\sigma+2\lambda
G_{r}\coth\frac{d_{N}}{\lambda}}\left(  \frac{\cosh\frac{z-d_{N}}{\lambda}%
}{\sinh\frac{d_{N}}{\lambda}}+\frac{\cosh\frac{z}{\lambda}}{\sinh\frac{d_{N}%
}{\lambda}}\right)  \cos\alpha\sin\alpha,\\
\frac{\mu_{sy}(z)}{\mu_{s}^{0}} &  =-\frac{\sinh\frac{2z-d_{N}}{2\lambda}%
}{\sinh\frac{d_{N}}{2\lambda}}-\frac{2\lambda G_{r}}{\sigma+2\lambda
G_{r}\coth\frac{d_{N}}{\lambda}}\left(  \frac{\cosh\frac{z-d_{N}}{\lambda}%
}{\sinh\frac{d_{N}}{\lambda}}\cos^{2}\alpha-\frac{\cosh\frac{z}{\lambda}%
}{\sinh\frac{d_{N}}{\lambda}}\sin^{2}\alpha\right)  ,\\
\mu_{sz}(z) &  =0,
\end{align}
which leads to the components of spin current normal to the interfaces
\begin{align}
\frac{j_{sx}(z)}{j_{s0}^{\mathrm{SH}}} &  =-\frac{2\lambda G_{r}\tanh
\frac{d_{N}}{2\lambda}}{\sigma+2\lambda G_{r}\coth\frac{d_{N}}{\lambda}%
}\left(  \frac{\sinh\frac{z-d_{N}}{\lambda}}{\sinh\frac{d_{N}}{\lambda}}%
+\frac{\sinh\frac{z}{\lambda}}{\sinh\frac{d_{N}}{\lambda}}\right)  \cos
\alpha\sin\alpha,\\
\frac{j_{sy}(z)}{j_{s0}^{\mathrm{SH}}} &  =\frac{\cosh\frac{2z-d_{N}}%
{2\lambda}-\cosh\frac{d_{N}}{2\lambda}}{\cosh\frac{d_{N}}{2\lambda}}%
+\frac{2\lambda G_{r}\tanh\frac{d_{N}}{2\lambda}}{\sigma+2\lambda G_{r}%
\coth\frac{d_{N}}{\lambda}}\left(  \frac{\sinh\frac{z-d_{N}}{\lambda}}%
{\sinh\frac{d_{N}}{\lambda}}\cos^{2}\alpha-\frac{\sinh\frac{z}{\lambda}}%
{\sinh\frac{d_{N}}{\lambda}}\sin^{2}\alpha\right)  .
\end{align}
The total current is the sum of those from the two ferromagnets at the top and
bottom; in contrast to the parallel $\hat{m}=\pm\hat{m}^{\prime}$ configuration, they do not
feel each other. We can extend the discussion from the previous subsection:
the second F can be a spin current source, and we can switch this source on by
rotating the magnetization from perpendicular to (anti)parallel configuration.

The longitudinal and transverse electric currents read
\begin{align}
\frac{j_{c,\mathrm{long}}(z)}{j_{c}^{0}} &  =1+\theta_{\mathrm{SH}}^{2}%
\frac{\cosh\frac{2z-d_{N}}{2\lambda}}{\cosh\frac{d_{N}}{2\lambda}}%
+\theta_{\mathrm{SH}}^{2}\frac{2\lambda G_{r}\tanh\frac{d_{N}}{2\lambda}%
}{\sigma+2\lambda G_{r}\coth\frac{d_{N}}{\lambda}}\left(  \frac{\sinh
\frac{z-d_{N}}{\lambda}}{\sinh\frac{d_{N}}{\lambda}}\cos^{2}\alpha-\frac
{\sinh\frac{z}{\lambda}}{\sinh\frac{d_{N}}{\lambda}}\sin^{2}\alpha\right)  ,\\
\frac{j_{c,\mathrm{trans}}(z)}{j_{c}^{0}} &  =\theta_{\mathrm{SH}}^{2}%
\frac{2\lambda G_{r}\tanh\frac{d_{N}}{2\lambda}}{\sigma+2\lambda G_{r}%
\coth\frac{d_{N}}{\lambda}}\left(  \frac{\sinh\frac{z-d_{N}}{\lambda}}%
{\sinh\frac{d_{N}}{\lambda}}+\frac{\sinh\frac{z}{\lambda}}{\sinh\frac{d_{N}%
}{\lambda}}\right)  \cos\alpha\sin\alpha.
\end{align}
Since the angle-dependent contributions vanish upon integration over $z$, there
is no magnetoresistance in the perpendicular configuration.

\subsection{Controlling the spin-transfer torque}

\begin{figure}[ptb]
\includegraphics[width=0.45\textwidth,angle=0]{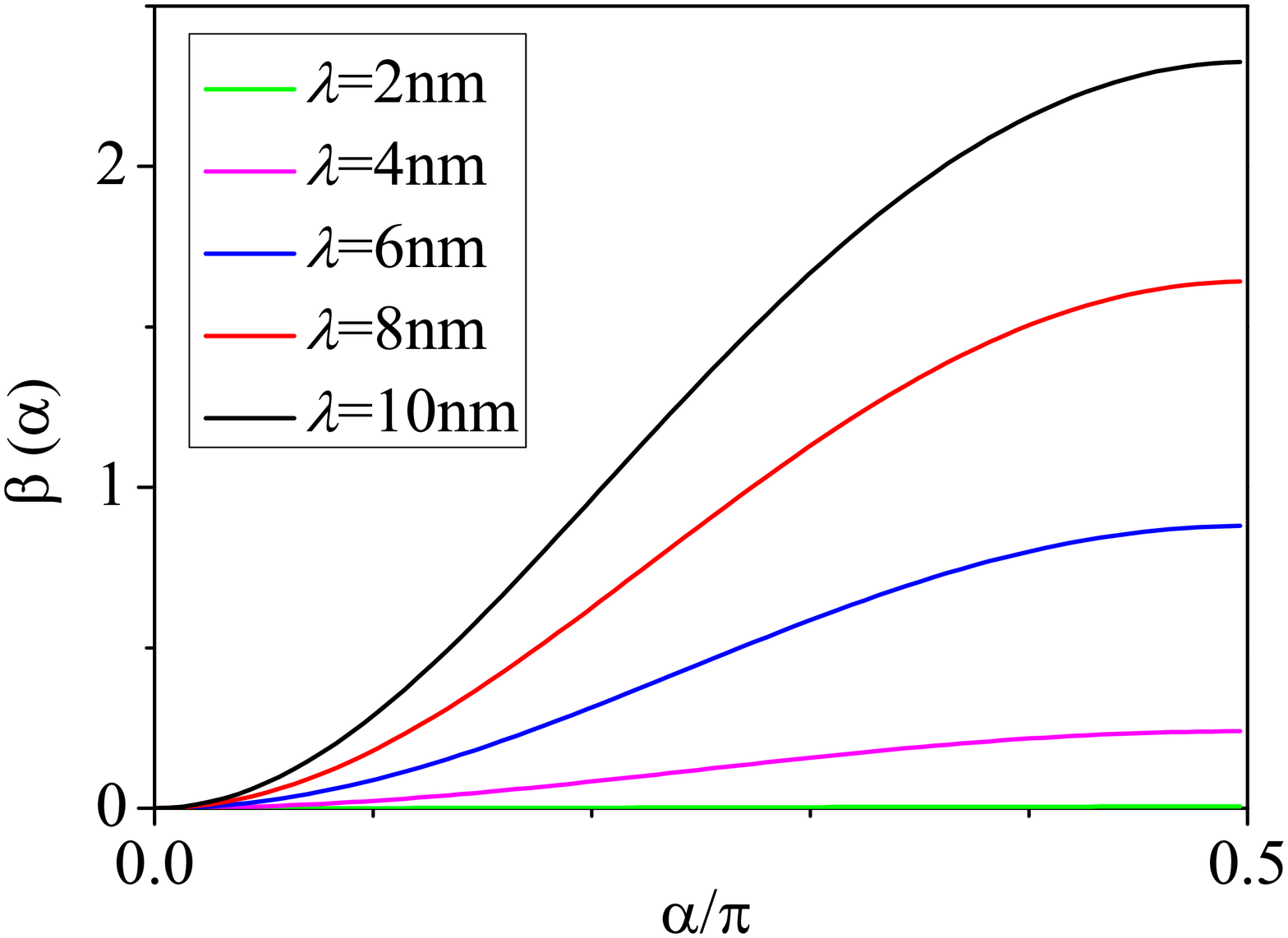}\caption{(Color
online) The ratio $\beta(\alpha)$ which characterize how $\vec{\tau
}_{\mathrm{stt}}^{\mathrm{(B)}}$ changes with respect to the relative
orientation between $\hat{m}$ and $\hat{m}^{\prime}$. We adopt the transport
parameters $d_{N}=12$~nm, $\rho=8.6\times10^{-7}\mathrm{\Omega\operatorname{m}%
}$, and $G_{r}=5\times10^{14}\mathrm{\Omega}^{-1}\mathrm{\operatorname{m}%
}^{-2}$.}%
\label{ratio-torque-change}%
\end{figure}

Like the SMR, the STT at the N$|$F interface depends on the relative
orientation between $\hat{m}$ and $\hat{m}^{\prime}$, too. We may pin $\hat
{m}=\hat{x}$ and observe how the STT at the bottom magnet, $\vec{\tau
}_{\mathrm{stt}}^{\mathrm{(B)}}\left(  \hat{m},\hat{m}^{\prime}\right)  $,
changes with rotating $\hat{m}^{\prime}=\hat{x}\cos\alpha+\hat{y}\sin\alpha$.
Figure~\ref{ratio-torque-change} displays the ratio $\beta$ defined as
\begin{equation}
\beta(\alpha)\equiv\frac{\left\vert \vec{\tau}_{\mathrm{stt}}^{\mathrm{(B)}%
}(\hat{x},\hat{x})-\vec{\tau}_{\mathrm{stt}}^{\mathrm{(B)}}(\hat{x},\hat
{x}\cos\alpha+\hat{y}\sin\alpha)\right\vert }{\left\vert \vec{\tau
}_{\mathrm{stt}}^{\mathrm{(B)}}(\hat{x},\hat{x})\right\vert },
\end{equation}
as a function of $\alpha$ for some spin-diffusion lengths. Only when $\lambda\ll
d_{N}$, $\beta$ remains constant under rotation of $\hat{m}^{\prime}$. A
larger spin-mixing conductance and smaller $d_{N}$ enhances the SMR as well as
angle dependence of $\beta$. This modification of the STT should lead to
complex dynamics of the spin valve in the presence of an applied current and
will be the subject of a subsequent study.

\section{Summary}

\label{summary} We developed a theory for the SMR in N$|$F and F$|$N$|$F
systems that takes into account the spin-orbit coupling in N as well as the
spin-transfer at the N$|$F interface(s). In a N$|$F bilayer system, the SMR
requires spin-flip in N and spin-transfer at the N$|$F interface. Our results
explain the SMR measured in Ref.~\onlinecite{Nakayama12} both qualitatively
and quantitatively with transport parameters that are consistent with other
experiments. The degrees of spin accumulation in N that can be controlled by
the magnetization direction is found to be very significant. In the presence
of an imaginary part of the spin-mixing conductance $G_{i}$ we predicted a
AHE-like signal (SHAHE). Such a signal was observed in
Ref.~\onlinecite{Althammer12} and can be explained with values of $G_{i}$ that
agree with first principles calculations.\cite{Jia11} We furthermore analyzed
F$|$N$|$F spin valves for parallel and perpendicular magnetization
configurations. A maximal SMR $\sim\theta_{\mathrm{SH}}^{2}$ is found for a
collinear magnetization configuration in the limit that the spin-diffusion length
is much larger than the thickness of the normal spacer. The SMR vanishes when
rotating the two magnetizations into a fixed perpendicular constellation. The
SMR torques under applied currents in N are expected to lead to magnetization
dynamics of N$|$F and F$|$N$|$F structures.

\begin{acknowledgments}
This work was supported by FOM (Stichting voor Fundamenteel Onderzoek der
Materie), EU-ICT-7 \textquotedblleft{MACALO,}\textquotedblright\ the ICC-IMR,
DFG Priority Programme 1538 \textquotedblleft{Spin-Caloric Transport}%
\textquotedblright\ (GO 944/4), and KAKENHI (Grant-in-Aid for Scientific
Research) C (22540346).
\end{acknowledgments}


\begin{thebibliography}{99}                                                                                               %


\bibitem {Bader10}S. D. Bader and S. S. P. Parkin, Ann. Rev. Cond. Matt. Phys.
\textbf{1}, 71 (2010).

\bibitem {Sinova12}J. Sinova and I \v{Z}uti\'{c}, Nature Mater. \textbf{11},
368 (2012).

\bibitem {Jungwirth12}For a review see: T. Jungwirth, J. Wunderlich, and K.
Olejn\'{\i}k, Nature Mater. \textbf{11}, 382 (2012).

\bibitem {Ando08}K. Ando, S. Takahashi, K. Harii, K. Sasage, J. Ieda, S.
Maekawa, and E. Saitoh, Phys. Rev. Lett. \textbf{101}, 036601 (2008).

\bibitem {Miron11}I. M. Miron, K. Garello, G. Gaudin, P.-J. Zermatten, M. V.
Costache, S. Auffret, S. Bandiera, B. Rodmacq, A. Schuhl, and P. Gambardella,
Nature \textbf{476}, 189 (2011).

\bibitem {Liu12}L. Liu, C. F. Pai, Y. Li, H. W. Tseng, D. C. Ralph, and R. A.
Buhrman, Science \textbf{336}, 555 (2012).

\bibitem {Saitoh06}E. Saitoh, M. Ueda, H. Miyajima, and G. Tatara, Appl. Phys.
Lett. \textbf{88}, 182509 (2006).

\bibitem {Mosendz10-1}O. Mosendz, J. E. Pearson, F. Y. Fradin, G. E. W. Bauer,
S. D. Bader, and A. Hoffmann, Phys. Rev. Lett. \textbf{104}, 046601 (2010).

\bibitem {Mosendz10-2}O. Mosendz, V. Vlaminck, J. E. Pearson, F. Y. Fradin, G.
E. W. Bauer, S. D. Bader, and A. Hoffmann, Phys. Rev. B \textbf{82}, 214403 (2010).

\bibitem {Czeschka11}F. D. Czeschka, L. Dreher, M. S. Brandt, M. Weiler, M.
Althammer, I.-M. Imort, G. Reiss, A. Thomas, W. Schoch, W. Limmer, H. Huebl,
R. Gross, and S. T. B. Goennenwein, Phys. Lett. Rev. \textbf{107}, 046601 (2011).

\bibitem {Uchida08}K. Uchida, S. Takahashi, K. Harii, J. Ieda, W. Koshibae, K.
Ando, S. Maekawa, and E. Saitoh, Nature \textbf{455}, 778 (2008).

\bibitem {Jaworski10}C. M. Jaworski, J. Yang, S. Mack, D. D. Awschalom, J. P.
Heremans, and R. C. Myers, Nature Mater. \textbf{9}, 898 (2010).

\bibitem {Uchida10}K. Uchida, J. Xiao, H. Adachi, J. Ohe, S. Takahashi, J.
Ieda, T. Ota, Y. Kajiwara, H. Umezawa, H. Kawai, G. E.W. Bauer, S. Maekawa,
and E. Saitoh, Nature Mater. \textbf{9}, 894 (2010).

\bibitem {Weiler12}M. Weiler, M. Althammer, F. D. Czeschka, H. Huebl, M. S.
Wagner, M. Opel, I.-M. Imort, G. Reiss, A. Thomas, R. Gross, and S. T. B.
Goennenwein, Phys. Rev. Lett. \textbf{108}, 106602 (2012).

\bibitem {Kajiwara10}Y. Kajiwara, K. Harii, S. Takahashi, J. Ohe, K. Uchida,
M. Mizuguchi, H. Umezawa, H. Kawai, K. Ando, K. Takanashi, S. Maekawa, and E.
Saitoh, Nature \textbf{464}, 262 (2010).

\bibitem {Brataas00}A. Brataas, Yu. V. Nazarov, and G. E. W. Bauer, Phys. Rev.
Lett. \textbf{84}, 2481 (2000); Eur. Phys. J. B \textbf{22}, 99 (2001).

\bibitem {Jia11}X. Jia, K. Liu, K. Xia, and G. E. W. Bauer, Eurphys. Lett.
\textbf{96}, 17005 (2011).

\bibitem {Heinrich11}C. Burrowes, B. Heinrich, B. Kardasz, E. A. Montoya, E.
Girt, Y. Sun, Y. Y. Song, and M. Wu, Appl. Phys. Lett. \textbf{100}, 092403 (2012).

\bibitem {Ashcroft76}N. W. Ashcroft and N. D. Mermin, Solid State Physics
(Saunders, Philadelphia, 1976).

\bibitem {McGuire75}T. R. McGuire and R. I. Potter, IEEE Trans. Magn.
\textbf{MAG-11}, 1018 (1975).

\bibitem {Thompson75}D. A. Thompson, L. T. Romankiw, and A. F. Mayadas, IEEE
Trans. Magn. \textbf{MAG-11}, 1039 (1975).

\bibitem {Nagaosa10}N. Nagaosa, J. Sinova, S. Onoda, A. H. MacDonald, and N.
P. Ong, Rev. Mod. Phys. \textbf{82}, 1539 (2010).

\bibitem {Fert08}A. Fert, Rev. Mod. Phys. \textbf{80}, 1517 (2008).

\bibitem {Huang12}S. Y. Huang, X. Fan, D. Qu, Y. P. Chen, W. G. Wang, J. Wu,
T. Y. Chen, J. Q. Xiao, and C. L. Chien, Phys. Rev. Lett. \textbf{109}, 107204 (2012).

\bibitem {Nakayama12}H. Nakayama, M. Althammer, Y.-T. Chen, K. Uchida, Y.
Kajiwara, D. Kikuchi, T. Ohtani, S. Gepr\"{a}gs, M. Opel, S. Takahashi, R.
Gross, G. E. W. Bauer, S. T. B. Goennenwein, and E. Saitoh, arXiv:1211.0098 (cond-mat.mes-hall).

\bibitem {Dyakonov07}M. I. Dyakonov, Phys. Rev. Lett. \textbf{99}, 126601 (2007).

\bibitem {Takahashi06}S. Takahashi, H. Imamura, and S. Maekawa, in
\textit{Concepts in Spin Electronics}, edited by S. Maekawa (Oxford University
Press, U.K., 2006), pp. 343-370.

\bibitem {Valet93}T. Valet and A. Fert, Phys. Rev. B \textbf{48}, 7099 (1993).

\bibitem {Liu11}L. Liu, R. A. Buhrman, and D. C. Ralph, arXiv:1111.3702 (cond-mat.mes-hall).

\bibitem {Althammer12}M. Althammer \textit{et al.}, unpublished.

\bibitem {Schep97}K. M. Schep, J. B. A. N. van Hoof, P. J. Kelly, G. E. W.
Bauer, and J. E. Inglesfield, Phys. Rev. B \textbf{56}, 10805 (1997).
\end{thebibliography}
\end{document}